\documentclass[11pt]{article}
\usepackage[a4paper,hmargin=1.0in,vmargin=1.0in]{geometry}
\usepackage{amsmath,amsthm,amssymb,setspace,sectsty,bbm,graphicx,mathtools}
\usepackage[round]{natbib}
\usepackage[usenames,dvipsnames]{xcolor}
\usepackage[linktocpage=true,pagebackref=true,colorlinks,allcolors=blue,bookmarks=true,bookmarksopen,bookmarksnumbered]{hyperref}

\usepackage{booktabs}
\usepackage{pdfpages}
\newtheoremstyle{DStheorem}
{\topsep}
{\topsep}
{\itshape}
{0pt}
{\scshape}
{.}
{ }
{\thmname{#1}\thmnumber{ #2}\thmnote{ (#3)}}
\theoremstyle{DStheorem}
\newtheorem{theorem}{Theorem}[section]
\newtheorem{lemma}[theorem]{Lemma}
\newtheorem{claim}[theorem]{Claim}
\newtheorem{corollary}[theorem]{Corollary}

\let\oldproofname=\proofname
\renewcommand{\proofname}{\rm\sc{\oldproofname}}

\newcommand{\MyAbove}[2]{\genfrac{}{}{0pt}{}{#1}{#2}}

\newcommand{\eps}{\epsilon}
\newcommand{\indicatorx}{1}
\newcommand{\pr}[1]{{\rm Pr} \left[ #1 \right]}
\newcommand{\prpar}[1]{{\rm Pr} [ #1 ]}
\newcommand{\prsub}[2]{{\rm Pr}_{#1} \left[ #2 \right]}
\newcommand{\prparsub}[2]{{\rm Pr}_{#1} [ #2 ]}
\newcommand{\ex}[1]{{\mathbb E} \left[ #1 \right]}
\newcommand{\expar}[1]{{\mathbb E} [ #1 ]}
\newcommand{\exsub}[2]{{\mathbb E}_{#1} \left[ #2 \right]}
\newcommand{\exparsub}[2]{{\mathbb E}_{#1} [ #2 ]}
\newcommand{\exsubpar}[2]{{\mathbb E}_{#1} [ #2 ]}
\newcommand{\opt}{\mathrm{OPT}}
\newcommand{\optcust}{\opt^{\mathrm{cust}}}
\newcommand{\optinc}{\opt^{\mathrm{inc}}}
\newcommand{\pic}{\pi^{(c)}}

\newcommand{\rev}{{\cal R}}

\newcommand{\bbR}{\mathbb{R}}
\newcommand{\mybern}{\mathrm{Bernoulli}}

\newcommand{\ignore}[1]{}
\newcommand{\customeqref}[2]{(\hyperref[#1]{#2})}
\newcommand{\revcustomized}{\mathcal{R}^{\text{\sf cust}}}
\newcommand{\revinclusive}{\mathcal{R}^{\text{\sf inc}}}
\newcommand{\finclusive}{f^{\text{\sf inc}}}
\newcommand{\fcustomized}{f^{\text{\sf cust}}}

\def\C{{\mathcal C}}
\def\S{{\mathcal S}}
\def\E{{\mathcal E}}

\def\revcustP{\mathcal{R}^{\mathrm{cust}}_{P}}
\def\revincP{\mathcal{R}^{\mathrm{inc}}_{P}}

\newcommand{\median}{\mathrm{median}}
\newcommand{\cust}{\mathrm{cust}}
\newcommand{\inc}{\mathrm{inc}}

\newcommand{\algcust}{\mathrm{ALG}_{\cust}}
\newcommand{\algcustobv}{\mathrm{ALG}^{\mathrm{obv}}_{\cust}}
\newcommand{\alginc}{\mathrm{ALG}_{\inc}}

\newcommand{\lpcust}{\mathrm{LP}_{\cust}}

\newcommand{\algatar}{\mathrm{ALG}_{\mathrm{AE25}}}
\newcommand{\lpatar}{\mathrm{LP}_{\mathrm{AE25}}}
\newcommand{\greedycust}{\mathrm{Greedy}_{\cust}}
\newcommand{\greedyinc}{\mathrm{Greedy}_{\inc}}
\newcommand{\fproxy}{f^{\text{\sf proxy}}}
\newcommand{\revproxy}{{\cal R}^{\text{\sf proxy}}}
\newcommand{\X}{\mathcal{X}}
\newcommand{\mycust}{\text{\sf cust}}
\newcommand{\myinc}{\text{\sf inc}}
\newcommand{\myproxy}{\text{\sf proxy}}
\newcommand{\sal}{\textsf{sal}}
\newcommand{\ratg}{\textsf{rating}}
\newcommand{\rem}{\textsf{rem}}
\newcommand{\lev}{\textsf{level}}
\newcommand{\skl}{\textsf{skill}}
\newcommand{\qual}{\textsf{qual}}
\newcommand{\applies}{\textsf{applies}}
\newcommand{\views}{\textsf{views}}

\definecolor{mysilver}{RGB}{220,220,220}

\newcounter{segevcounter}

\sectionfont{\large} \subsectionfont{\normalsize}
\allowdisplaybreaks
\makeindex
\newcommand{\changelocaltocdepth}[1]{%
  \addtocontents{toc}{\protect\setcounter{tocdepth}{#1}}%
  \setcounter{tocdepth}{#1}%
}

\begin{document}

\begin{titlepage}

\title{Revenue Maximization in Choice-Based Matching Markets}

\author{%
Dan Nissim\thanks{School of Mathematical Sciences, Tel Aviv University, Tel Aviv 69978, Israel. Email: {\tt nissim.dan@gmail.com}.}%
\and Danny Segev\thanks{School of Mathematical Sciences and Coller School of Management, Tel Aviv University, Tel Aviv 69978, Israel. Email: {\tt segevdanny@tauex.tau.ac.il}. Supported by Israel Science Foundation grant 1407/20.}%
\and Alfredo Torrico\thanks{ Daniel J. Epstein Department of Industrial and Systems Engineering, University of Southern California, Los Angeles, CA 90089, United States. Email: {\tt atorrico@usc.edu}.}}

\date{}
\maketitle

\setcounter{page}{200}
\thispagestyle{empty}

\begin{abstract}
The primary contribution of this paper resides in devising constant-factor approximation guarantees for revenue maximization in two-sided matching markets with general pairwise rewards. A major distinction between our work and state-of-the-art results in this context \citep{AshlagiKMSS22, TorricoCL22} is that, for the first time, we are able to address \emph{reward maximization}, reflected by assigning each customer-supplier pair an arbitrarily-valued reward. Our performance guarantees apply to two different settings: the  \emph{customized} model and the \emph{inclusive} model.
The fundamental difference between these settings lies in whether the platform should display to each supplier all selecting customers, as in the inclusive model, or whether the platform can further personalize this set, as in the customized model. Technically speaking, our algorithmic approaches and their analysis revolve around presenting a closely related combinatorial problem, a novel linear relaxation, and developing tailor-made analytical ideas.
In both models considered, these ingredients allow us to overcome the lack of submodularity and subadditivity that stems from pairwise rewards, plaguing the applicability of existing methods. We complement our theoretical results with an extensive computational study under different types of instances where we analyze the empirical revenue and running time performance of our proposed methods.
\end{abstract}

    
\end{titlepage}

\setcounter{page}{200}
\pagestyle{empty}
\tableofcontents

\newpage
\pagestyle{plain}
\setcounter{page}{1}

\section{Introduction}\label{sec:intro}

Choice-based matching platforms such as Tinder, Upwork and BlaBlaCar have become integral parts of our daily lives, providing essential solutions for a wide range of services, including  transportation, accommodation, dating, and job searching, to name a few. Platforms of this nature have significantly lowered the entry barriers for the parties they connect, with service suppliers on one side and customers on the other, leading to the present-day evolution of two-sided markets. One of the main features inherent to  these platforms is ``decentralization'', since customer-supplier pairs are not centrally dictated but rather result from their personal choices. Consequently, platforms serve as mediators that  carefully craft the set of displayed alternatives from which customers make their selections. 
A recent trend in this context has been investigating questions around analyzing different forms of interactions and interventions to improve pivotal performance metrics, prompting the development of new modeling frameworks and algorithmic solutions.
 
 \paragraph{Foundational algorithmic work in choice-based matching markets.} In the specific context of two-sided assortment optimization, let us discuss three recent papers that laid the groundwork for exploring the interplay between choice-based matching markets and computational revenue management. 
For ease of exposition, we proceed by presenting a succinct overview of these settings, focusing on high-level concepts.
Later on, their complete mathematical description will be provided in Section~\ref{subsec:description_of_the_problem}, followed by an outline of directly related work and further background in Section~\ref{subsec:previous_work}.

To our knowledge, the foundational work of \cite{AshlagiKMSS22} pioneered the incorporation of assortment optimization into two-sided markets, exploring nuanced challenges in this context. 
At a high level, the authors introduced a modeling framework where we play the role of a platform that acts as an intermediary between customers and suppliers, forming the two sides of the matching market. 
In a nutshell, the platform offers each customer a personalized set of suppliers. With respect to these so-called ``menus'', customers simultaneously and independently  employ an individual choice model either to select a supplier from their menu or to opt for the outside option, i.e., remain unmatched. 
In turn, suppliers review their random set of selecting customers and decide, again according to an individual choice model, whether to engage with one of these customers or to choose the outside option. 
Consequently, a customer-supplier match is successful only when both parties select each other.
In this setting, the platform wishes to determine personalized menus that would maximize the creation of as many matches as possible in expectation.

Subsequently,~\cite{TorricoCL22}  significantly expanded the scope of this framework by incorporating two key features.
First, rather than focusing on the expected number of matches, they considered their quality by introducing supplier-related rewards, aiming to maximize the total expected reward.
Second, the authors allowed greater diversity among customers and suppliers, moving away from the very specific choice model considered by~\cite{AshlagiKMSS22}.
The third paper along these lines is that of~\cite{AhmedSB22}, who further generalized the objective function by studying reward maximization in settings where each customer-supplier pair has a designated reward, realized when they are eventually matched.
In other words, rather than assuming
that rewards are solely attributed to suppliers, their model features generally-structured pairwise rewards. That said, \cite{AhmedSB22} do not consider a sequential matching process, and instead focus on a problem formulation where customers and suppliers make simultaneous matching decisions.

\paragraph{Informal contributions.} 
The current paper provides constant-factor approximation guarantees for reward maximization in two-sided assortment optimization problems. A major distinction between our work and that of~\cite{AshlagiKMSS22} and~\cite{TorricoCL22} is that, for the first time, we focus on \emph{reward maximization}, reflected by assigning each customer-supplier pair an arbitrarily-structured reward. These results are derived for our newly introduced \emph{customized} model as well as for the \emph{inclusive} model.
The fundamental difference between these models lies in whether the platform should display to each supplier all selecting customers, as in the inclusive model, or whether the platform can tailor-make this set, as in the customized model. {From a practical standpoint, both models have their own trade-offs. In the inclusive model, suppliers have more flexibility to choose among potential customers, however, at the cost of reducing the platform's revenue since the supplier may choose a more preferred customer who yields a smaller reward. On the other hand, in the customized model, the platform personalizes the set of customers for its revenue benefit, but this approach may remove more appealing customers from the supplier's choice set. Despite these differences, both models arise in practice as in the end suppliers ultimately see all interested customers, but some are prioritized and listed higher to help the user navigate the platform and reduce fatigue.
In the course of investigating} these models, we develop novel algorithmic tools and analytical ideas to deal with quite a few technical hurdles such as non-submodularity and non-monotonicity of our reward functions. {We complement these technical results with a detailed computational study where we assess the performance of our proposed algorithms across different classes of instances.}

\subsection{Model description: Sequential two-sided matching}\label{subsec:description_of_the_problem}

\paragraph{High-level setting.} 
An instructive way of viewing our models of interest is through a bipartite graph, with the set of customers $\C$
on one side and with the set of suppliers $\S$ on the other. 
At a high level, the solution concept we are studying consists of deciding on a so-called menu $M = (M_i)_{i \in \C}$, in which $M_i \subseteq \S$ represents the set (or assortment) of suppliers offered to each customer $i \in \C$; the latter can be equivalently thought of as picking the set of edges adjacent to that customer. Given this menu, a random matching in its induced graph will be sequentially constructed via a two-step process, with actions taken by both parties involved, customers and suppliers.

As explained below, each customer $i \in \C$ initially selects at most one supplier out of her menu $M_i$, guided by an individual Multinomial Logit (MNL) choice model.
In the second stage, each supplier makes a matching decision, again according to an individual MNL choice model, with the assortment of customers shown depending on the model of interest. Specifically, in the inclusive model, each supplier observes all customers who have chosen her. On the other hand, in the customized model, the platform is allowed to filter the set of customers who have chosen each supplier.
In contrast to~\cite{AshlagiKMSS22} and~\cite{TorricoCL22}, we adopt the general viewpoint  of~\cite{AhmedSB22} and focus on maximizing \emph{pairwise rewards}, where $r_{ij}$ represents the reward garnered when customer $i \in \C$ and supplier $j \in \S$ select each other.
This mutual selection occurs when the edge $(i, j)$ appears in the resulting customer-supplier matching.
In what follows, we provide the finer details of these models.

\paragraph{Step 1: Customer choices.} Whether one considers the customized model or the inclusive model, the first step of our matching-creation process is identical. Here, for a given menu $M$, each customer $i \in \C$ independently selects from her set of suppliers $M_i$ via an individual MNL model, corresponding to the random selection of at most one adjacent edge. 
For convenience, the outside option in this context will be denoted by ``supplier 0''.
To capture the above-mentioned choice, we make use of $\{ u_{ij} \}_{j \in \S}$ to designate the preference weights customer $i$ associates with the various suppliers, assuming by convention that the outside option's weight is normalized to $u_{i0} = 1$.
As such, customer $i$ decides to select each supplier $j \in M_i$ with probability
\[ {\pi_i}(j, M_i) ~~=~~ \frac{u_{ij}}{1 + \sum_{k \in M_i} u_{ik}} \ .\]
Alternatively, this customer selects the outside option with probability
\[\pi_i(0, M_i) ~~=~~ \frac{1}{1 + \sum_{k \in M_i} u_{ik}} \ .\]
Once all customers have made their decisions, we use $C^M_j \subseteq \C$ to denote the random set of customers who picked supplier $j \in \S$.
That is, letting $I^M_{ij}$ be a binary random variable that indicates whether customer $i$ picked supplier $j$ or not, we have $C^M_j = \{ i \in \C : I^M_{ij} = 1 \}$. 
Now, the second stage of our matching process depends on whether we consider the customized or the inclusive model, as further explained in the upcoming paragraphs.

\paragraph{Step 2 of the customized model: Supplier choices.} 
To elaborate on how things evolve in the customized model, suppose that $C_j$ is a realization of the random set $C^M_j$ of customers who picked supplier $j \in \S$ upon completing step~1. 
Then, the choice of supplier $j$ will be governed by an individual MNL model, in which $\{ w_{ij} \}_{i \in \C}$ stand for the preference weights associated with the underlying collection of customers.
In the current model, customization is captured by a personalized set of customers $T_j \subseteq C_j$ that will be presented to supplier $j$, who will only select between the alternatives in $T_j$.
The platform customizes $T_j$ for its own benefit, meaning that $T_j$ is determined to maximize the expected reward gained from supplier $j$. We remark that $T_j$ may very well be a proper subset of $C_j$, since the MNL expected reward is not necessarily monotone. Based on these considerations, letting {$\fcustomized_j(C_j)$} be our expected reward due to supplier $j$, we have 
\begin{equation} \label{eqn:f_customized_definition}
    \fcustomized_j(C_j) ~~=~~ \max_{T_j \subseteq C_j} \left\{ \sum_{i \in T_j} r_{ij} \cdot \frac{w_{ij}}{1 + \sum_{\ell \in T_j} w_{\ell j}} \right\} \ .
\end{equation}
{
The customized model focuses on personalization which help minimize user fatigue and  increase the platform's revenue. We emphasize that models with customization have been previously considered in the revenue management literature, see, e.g.,~\citep{el2023joint}. Even though we interpret $r_{ij}$'s as rewards, each of those values can represent the utility or welfare of matching $i$ with $j$; a more natural concept in two-sided markets such as dating platforms~\citep{RiosT23} which do filter the set of displayed options.
}

\paragraph{Step 2 of the inclusive model: Supplier choices.} Moving on to consider the inclusive case, we assume that the platform cannot filter out selecting customers. 
In other words, each supplier $j \in S$ will be picking out of all customers who selected her during step~1, i.e., out of the entire set $C_j$.
As such, letting {$\finclusive_j(C_j)$}
be the expected reward gained from supplier $j$, we have 
\begin{equation} \label{eqn:f_inclusive_definition}
\finclusive_j(C_j) ~~=~~ \sum_{i \in C_j} r_{ij} \cdot \frac{w_{ij}}{1 + \sum_{\ell \in C_j} w_{\ell j}} \ .
\end{equation}
{As we mentioned earlier, this model provides greater flexibility to the supplier as the pool of customers is potentially larger, thereby incentivizing the supplier to stay in the platform. However, the expected inclusive reward is at most the customized reward, since for any $C_j\subseteq\C$, we have $\fcustomized_j(C_j)\geq \finclusive_j(C_j)$.}

\paragraph{Objective.} Viewing the random process discussed above from a graph-theoretic standpoint, the selections made during step~1 guarantee that each customer has at most one adjacent edge.
Then, subsequent choices made during step~2 ensure that each supplier has at most one adjacent edge. 
In other words, we have just defined a random customer-supplier matching.
In the reward maximization problem, we wish to compute a menu $M = (M_i)_{i \in \C}$ whose expected reward $\rev( M )$  is maximized. The latter measure is specified by
\begin{equation} \label{eqn:definition_reward}
\rev(M) ~~=~~ \sum_{j \in S} {\ex{f_j(C^M_j)}} \ ,
\end{equation}
where the expectation above is taken over the randomness in $( C_j^M )_{j \in \S}$. 
In this representation, our random reward is decomposed into suppliers' contributions, and {when necessary, we distinguish both cases by writing $\revcustomized$ for $f_j=\fcustomized_j$ and $\revinclusive$ for $f_j=\finclusive_j$.}

\subsection{Previous work and open questions}\label{subsec:previous_work}

\paragraph{Introductory work.} 
To our knowledge,~\cite{AshlagiKMSS22} were the first to model assortment optimization in two-sided matching markets of the nature described above. 
Their paper focuses on a highly structured special case of the models described in Section~\ref{subsec:description_of_the_problem}, specifically assuming that: 
\begin{itemize}
    \item All customers make decisions according to the same MNL choice model.

    \item From a supplier's perspective, all customers have identical preference weights, whereas the outside option has a supplier-dependent weight.

    \item All edge rewards are identical.
\end{itemize}
We mention in passing that, with uniform rewards, the platform cannot increase revenues by filtering out customers, implying that the customized and inclusive models coincide.
The first technical contribution of~\cite{AshlagiKMSS22} was to prove that this specific model is strongly NP-hard via a reduction from the 3-partition problem.
On the positive side, they devised a constant-factor approximation, showing how to compute a menu whose expected number of matches is within factor $0.5 \cdot 10^{-5}$ of optimal. Their approach is based on separately considering two regimes, one with ``high-value'' suppliers ($u_j > 1$), and the other with ``low-value'' suppliers ($u_j \leq 1$). In the low-value regime, suppliers are aggregated into buckets, each containing suppliers whose preference weights are of similar magnitudes, so that suppliers within a given bucket can be treated as being essentially indistinguishable. Subsequently, \cite{AshlagiKMSS22} formulated a linear relaxation, where we wish to decide on the fractional number of suppliers from each bucket to be included in each customer's menu. An optimal solution to this relaxation is rounded, with suppliers in each bucket evenly distributed across customers’ menus to avoid excessive overlaps, ensuring that suppliers belonging to the same bucket are offered to roughly the same number of customers. In contrast, the high-value regime is approached by considering a relaxation where customers' outside options are ignored, thereby making single-supplier menus much more rewarding. Interestingly, this relaxation can be approximated within a constant factor.

\paragraph{Greater generality and improved approximations.} Subsequently, the work of \cite{TorricoCL22} studied the incorporation of two fundamental features. 
First, rather than focusing on maximizing the expected number of matches, the authors considered {\em supplier-specific} rewards.
Second, \cite{TorricoCL22} introduced variability among customers and suppliers by allowing general choice models for both parties;
that said, a crucial assumption is that the suppliers' choice model has a monotone and submodular demand function.
Under this assumption, they proved that the two-sided assortment optimization problem can be approximated within factor $1 - \frac{1}{e}$ of optimal, primarily relying on the continuous greedy algorithm for submodular maximization \citep{CalinescuCPV11}.
A particularly interesting aspect of this work is the investigation of the so-called multilinear relaxation, allowing one to incorporate structural constraints on feasible menus, dictated by business requirements such as limiting the number of profiles shown in dating apps. Once again, with supplier-specific rewards, the inclusive and customized models are equivalent, as the platform is incentivized to keep all customers who select a given supplier.

\paragraph{Customer-supplier rewards.}
Yet another relevant paper is that of \cite{AhmedSB22}, who were the first to consider pairwise rewards, albeit in a problem formulation where customers and suppliers are making concurrent matching decisions. Here, each customer-supplier pair is associated with its own reward, without any assumption on the relation between different pairs. 
The authors provided a mixed-integer linear formulation of this model and examined two greedy heuristics, illustrating that both may yield arbitrarily bad outcomes. 
Finally, \cite{AhmedSB22} suggested several relaxations for deriving parametric upper and lower bounds on the optimal reward, and conducted numerical experiments on synthetic data to evaluate the effectiveness of these relaxations.

\paragraph{Challenges.} Given this current state of affairs, we proceed by highlighting the most fundamental open questions related to revenue maximization in choice-based 
matching markets: 
\begin{itemize}
    \item \emph{Tractability of pairwise rewards.} 
    In comparison to the models studied by~\cite{AshlagiKMSS22} and~\cite{TorricoCL22}, for which constant-factor approximations are known, our models associate customer-supplier pairs with arbitrarily-structured rewards.
    To our knowledge, the algorithmic methods and analytical ideas proposed in prior research cannot be employed when considering pairwise rewards, since removing the assumption of uniform rewards or supplier-specific rewards takes away many useful properties. At present time, we are unaware of any non-trivial approximability result for general pairwise rewards.

    \item \emph{Dealing with lack of submodularity and subadditivity.} 
    Contrary to previous work, which exploits submodularity and monotonicity in one way or another, our models do not exhibit either of these properties. This discrepancy poses major challenges, since we can no longer leverage advancements in the realm of submodular maximization; see, e.g., \cite{NemhauserWF78}, \cite{CalinescuCPV11}, and \cite{BuchbinderF18Survey}.
    Elementary examples demonstrate {the non-submodularity of both $\finclusive_j$ and $\fcustomized_j$, and the non-monotonicity of the former, with generally structured customer-supplier rewards.}
    Moreover, our objective function {$\rev$} does not even satisfy subadditivity, a feature present in earlier papers \citep{AshlagiKMSS22, TorricoCL22}, allowing for reductions to more manageable subproblems.
\end{itemize}

\subsection{Main contributions}\label{subsec:main_contributions}
The primary contribution of this paper resides in devising efficient constant-factor approximations for revenue maximization in choice-based matching markets, under general pairwise rewards, for both the customized and inclusive models. A notable aspect of our approach is the sequential development of approximate structural simplifications: {starting from a proxy problem formulation whose structure is exploited to design a purely combinatorial method, and eventually refining some of these ideas further into a novel linear programming (LP) relaxation}.
In formulating this relaxation, rather than focusing on discrete menus, we translate our search space to appropriate downward-closed polyhedra of MNL-choice-induced probability vectors, showing how to losslessly migrate solutions between these two representations. Technically speaking, we derive these approximation guarantees by presenting novel bounding methods inspired by classic MNL-based assortment optimization, {leveraging combinatorial aspects of the proxy model,}
and developing tailor-made analytical ideas. One key insight that connects both approaches is the development of truncated solutions from auxiliary bounds, either from the proxy model or from the LP relaxation. To construct these truncated solutions we leverage the optimal value of a related per-customer assortment optimization problem which satisfy a crucial fixed point equation.
In both models considered, these ingredients allow us to overcome the lack of submodularity and subadditivity that stems from pairwise rewards, plaguing the applicability of existing algorithmic methods.
In the remainder of this section, we elaborate on the specifics of these contributions in greater detail.

\paragraph{Main result 1: Combinatorial approach.}
{As previously explained, the inclusive model is distinctively characterized by not allowing the platform to construct personalized assortments in step~2, in the sense that each supplier should be offered the entire set of customers who selected her in step~1. }
In Section~\ref{sec:purely-combinatorial}, we describe a combinatorial algorithm for efficiently computing a menu vector, guaranteeing that its expected inclusive reward is within a constant factor of optimal, as specified in the next theorem.

\begin{theorem} \label{thm:combinatorial} 
For the inclusive model, there exists an algorithm for computing a menu whose expected reward is within a factor $1/4$ of the optimal inclusive reward. 
\end{theorem}
Our approach uses a proxy matching model based on independent customer qualifications, whose expected reward upper-bounds that of the customized model and, consequently, the inclusive model. Moreover, this proxy optimization problem can be reformulated as monotone submodular maximization subject to a partition matroid constraint, allowing us to obtain a $\frac{ 1 }{ 2 }$-approximation via the greedy algorithm where each step can be efficiently implemented by solving an MNL-based assortment problem. To conclude our guarantee, we truncate the resulting menus using specific reward thresholds, retaining at least half of their expected proxy reward in the inclusive model.
 
\paragraph{Main result 2: An LP-based algorithm for both models.}
The customized model is distinctively characterized by allowing the platform to personalize suppliers' assortments, maximizing its revenue from a given set of customers. 
In Section~\ref{sec:lpbased_approach}, we present an LP-based algorithm for efficiently computing a fractional menu vector, ensuring that its expected inclusive reward is within a constant factor of the best possible expected customized reward, as formally stated in the following theorem.

\begin{theorem}\label{thm:main_algorithm}
For the inclusive model, there exists a randomized algorithm for computing a menu whose expected  reward is within a factor $\frac{ 1 }{ 2}$ of the optimal expected customized reward.
\end{theorem}

Our approach involves a novel linear relaxation of the customized model whose optimal solution yields a constant-factor approximation. This relaxation involves an extended MNL-choice polyhedron that simultaneously captures customers' and suppliers' choice preferences. 
At the same time, it dictates a linear relation between customers' and suppliers' MNL-choice probabilities, via appropriately-defined coefficients.
In this context, our analysis introduces a new bounding method where fractional solutions to the MNL-based linear formulation are migrated to a random environment. We then construct our final solution by appropriately truncating each component with specific supplier-reward thresholds obtained from the LP relaxation.

From Theorem~\ref{thm:main_algorithm}, we derive three related results. First, the guarantee also applies to the customized model, i.e., there exists a randomized algorithm for computing a menu whose expected customized reward is at least $1/2$ of optimal (Corollary~\ref{coro:main_customized}). This is possible since the customized model upper bounds the inclusive model. Second, one natural question in this context is whether one can obtain improved guarantees. Along these lines, we present a worst-case instance where the optimal expected customized reward is asymptotically $1/2$ of the optimal LP value (Lemma~\ref{lem:lowerbound_onehalf}). In other words, no LP-based approach using our relaxation can guarantee a constant factor better than $1/2$. Finally, we establish the \emph{reward gap} between the two models, defined as the worst-case ratio between the optimal expected customized reward and inclusive reward. In particular, we show that this gap is asymptotically 2 (Theorem~\ref{thm:revenue_gap}).

\paragraph{Computational study.}
We complement our theoretical analysis by empirically assessing the performance of our proposed methods under different types of instances. Specifically, we assess the reward guarantees and running time performance of our approximation algorithms as well as other benchmarks under synthetic instances. Then, using larger semi-synthetic instances, we analyze the reward performance under different levels of correlation between pairwise reward and  preference values. Our analysis shows that: (i) our methods run significantly faster than related benchmarks; (ii) their empirical expected reward performance is considerably better than the worst-case guarantees, and there is no significant difference between the customized and inclusive models, in contrast to the worst-case reward gap; (iii) regardless of the correlation level between reward and preference values, our reward-aware approaches perform significantly better than reward-oblivious methods; however, the gap narrows under positive and negative correlation, and widens under no correlation. In other words, reward-aware methods are most valuable when rewards and preferences are decoupled.

\subsection{Related work}\label{sec:related-work}

We conclude this section by succinctly reviewing three research threads that are directly related to our work. Specifically, we elaborate on the most relevant results in classic assortment optimization, describe recent advances in two-sided matching markets, and expand on the relation between our modeling framework and probe-commit stochastic matching.

\paragraph{Classic assortment optimization.} The vast majority of analytical work in assortment optimization has traditionally taken the perspective of a single customer, whose purchasing preferences are governed by a given choice model. Here, our prototypical objective is to identify a subset of products that maximizes the expected revenue attained due to the customer's selection. This setting was popularized by the pioneering work of  \citet{Talluri04} and has grown immensely since then in multiple directions, spanning offline, online, and dynamic models. Notably, \citet{Talluri04} proved that, when customer preferences are determined by Multinomial Logit choices, there exists an optimal unconstrained offer set which is revenue-ordered. Of particular interest are subsequent papers on this widely applicable model, studying the tractability of various structural  constraints \citep{RusmevichientongSS10,RusmevichientongSTT14,SumidaGRTD21,DesirGZ22}. Offline settings of this nature have also been analyzed under the Nested Logit model~\citep{DavisGT14,GallegoT14,LiRT15}, non-parametric preferences \citep{FariasJS13,AouadFRS18}, and the Markov chain based model \citep{BlanchetGG16,FeldmanT17,DesirGSY20}, to mention a few well-known examples. The specific choice model considered, along with extra side constraints, often lead to intractable settings, and therefore, much of the relevant literature is  focused on approximation algorithms \citep{RusmevichientongSTT14,AouadFRS18,DesirGSY20,DesirGZ22,Udwani25}. Concurrently, online assortment optimization has been an active research direction, and we refer avid readers to selected papers along these lines for additional discussion; see, e.g., \citep{GolrezaeiNR14,MaS20,GongGISUW22,FengNS22}.

\paragraph{Two-sided assortment optimization.} Beyond the papers we have thoroughly discussed in Section~\ref{subsec:previous_work} \citep{AshlagiKMSS22,TorricoCL22,AhmedSB22}, \cite{ElHousniHT24} have recently introduced a broader framework for two-sided assortment optimization, taking into account several classes of possible policies. On one extreme, they define \emph{fully static} policies, in which assortments are simultaneously displayed  to both sides of the market. On the other extreme, they consider \emph{fully adaptive} policies where, at each step, the platform displays an assortment to a single user, observes her choice, and possibly adapts future decisions.  
Within this framework, \cite{ElHousniHT24} studied the adaptivity gaps between different policy classes and provided a number of  approximation guarantees in this context.  
It is important to point out that the objective considered in this paper is maximizing the expected matching size, and we are not aware of any implications of their ideas to edge-dependent rewards.

\citet{AouadS23} introduced an online version of two-sided assortment optimization, where customers arrive in an online fashion, with the available suppliers varying from one stage to the next. At each arrival, the platform is asked to display an assortment of available suppliers to the incoming customer. In the adversarial setting, the customers' arrival and suppliers' availability can be arbitrarily chosen. Here, \citet{AouadS23} presented an online greedy algorithm that attains a competitive ratio of $\frac{1}{2}$, which is best possible. In addition, the authors attained improved competitive ratios for the i.i.d.\ setting and for specific customers' choice models. Similarly to previously-mentioned papers, their goal is once again maximizing the expected matching size, allowing one to exploit the submodularity of this function.

Motivated by dating platforms, \citet{RiosSZ23} and \cite{RiosT23} considered a dynamic model, where user preferences are determined by asymmetric edge-dependent probabilities, and where users can like/dislike multiple profiles in each step. The platform aims to maximize the expected matching size by displaying a subset of profiles to each arriving agent. In particular, \citet{RiosT23} presented an integer programming approach that guarantees a $(1- \frac{1}{e})$-approximation under different platform features. Our work substantially differs from these papers, which do not involve any choice modeling features.

\paragraph{Stochastic matching: The probe-commit model.} \cite{ChenIKMR09} studied stochastic matching in the so-called probe-commit model, also known as stochastic probing. Here, every edge $e$ of a general graph independently exists with probability $p_e$. In each step, the decision-maker probes a single edge; when it exists, this edge must be included in the final solution. The ultimate goal is to maximize the expected matching size. In the general version of this problem, each vertex is associated with a \emph{patience constraint}, limiting the number of incident edges that can be queried. \cite{ChenIKMR09} proposed a greedy algorithm that guarantees a $\frac{1}{4}$-approximation with respect to the optimal time-unbounded online algorithm; this result was subsequently improved to a $\frac{1}{2}$-approximation by  \cite{Adamczyk11}. Later on, \cite{BansalGLMNR12} presented an LP-based method for obtaining a $\frac{1}{4}$-approximation in general edge-weighted graphs and a $\frac{1}{3}$-approximation in bipartite graphs. Following a sequence improvements \citep{AdamczykGM15,BavejaCNSX18,BrubachGMS21}, the best known approximation factor for general graphs is $0.395$ \citep{PollnerRSW22}, with better guarantees for numerous special cases \citep{CostelloTT12, GamlathKS19, FuTWWZ21, PollnerRSW22, DerakhshanF23, BrubachGMS21, BorodinM25}. 
Our work departs from this literature since we simultaneously query a set of edges by deciding on customers' menus, and since the choice probabilities on both sides are endogenously determined by Multinomial Logit preferences.
\section{Purely Combinatorial Approach}
\label{sec:purely-combinatorial}

In what follows, we establish Theorem~\ref{thm:combinatorial} by developing a purely combinatorial polynomial-time algorithm that approximates the inclusive model within factor $\frac{ 1 }{ 4 }$. Toward this objective, in Section~\ref{subsec:upper_bound_comb}, we introduce a proxy matching model based on independent customer qualifications, whose expected reward upper-bounds that of the customized model and, consequently, the inclusive model. In Section~\ref{subsec:proxy_submodular}, we reformulate the proxy optimization as monotone submodular maximization subject to a partition matroid constraint, allowing us to obtain a $\frac{ 1 }{ 2 }$-approximation via the greedy algorithm. Here, even though the underlying ground set is exponentially large, it turns out that each greedy step can be efficiently implemented by recognizing it as an MNL-based assortment problem in disguise. Finally, in Section~\ref{subsec:comb-conversion}, we utilize supplier-specific reward thresholds to truncate the resulting menus, retaining at least half of their expected proxy reward in the inclusive model and thereby concluding our analysis.

\subsection{Upper-bounding mechanism}
\label{subsec:upper_bound_comb}

We begin by introducing an auxiliary two-sided matching process, referred to as the ``proxy model'', whose intended purpose is purely analytical. We show that its expected reward function provides an upper bound on its customized analogue, and therefore, also on the inclusive reward function. Moreover, we derive an explicit expression for each supplier's expected proxy reward, which will be useful in subsequent arguments.

\paragraph{The proxy model.} The setting we are about to present is identical to the customized and inclusive models in terms of input specification as well as in step~1 of our matching-creation process. Therefore, given a menu $M = (M_i)_{i \in \C}$, the upcoming discussion focuses on how step~2 of the proxy model operates in regard to supplier choices, once each customer $i \in \C$ has already selected from her set of suppliers $M_i$.

For this purpose, based on the terminology of Section~\ref{subsec:description_of_the_problem}, let us recall that $C^M_j = \{ i \in \C : I^M_{ij} = 1 \}$ stands for the random set of customers who picked supplier $j \in \S$ upon completing step~1. Here, $I^M_{ij}$ is a binary random variable that indicates whether customer $i$ picked supplier $j$ or not, meaning that $I^M_{ij} \sim \mybern( {\pi_i}(j, M_i) )$. In the proxy model, the choice of each supplier $j$ will not be governed by MNL preferences. Instead, independently of customer choices, this supplier draws a collection of mutually independent qualification indicators $\{ J_{ij} \}_{i \in \C}$, such that $J_{ij} \sim \mybern(\hat{w}_{ij})$ and $\hat{w}_{ij} = \frac{ w_{ij} }{ 1 + w_{ij} }$. Subsequently, out of all qualified customers who selected her during step~1, supplier $j$ picks the one whose reward is maximal. As such, letting $\fproxy_j(M)$ be the random reward collected by supplier $j$, we have
\begin{equation} \label{eqn:f_proxy_definition}
\fproxy_j(M) ~~=~~ \max_{i \in \C} \left\{ r_{ij} I_{ij}^M J_{ij} \right\} \ .
\end{equation}
Our objective is to compute a menu $M = (M_i)_{i \in \C}$ whose expected reward $\revproxy( M )$ is maximized. The latter measure is specified by
\[ \revproxy(M) ~~=~~ \sum_{j \in S} \ex{ \fproxy_j(M) } \ , \]
where the expectation above is taken over the randomness in $\{ I_{ij}^M \}_{i \in \C, j \in \S}$ and $\{ J_{ij} \}_{i \in \C, j \in \S}$.

\paragraph{Upper bound.} The following claim formalizes the upper-bounding property of the proxy model for every fixed menu. Its proof proceeds supplier by supplier, showing that, conditional on the set of customers who selected her, each supplier’s proxy reward stochastically dominates the reward generated by an optimal MNL assortment.

\begin{lemma}
\label{lem:proxy_UB}
$\revcustomized( M ) \leq \revproxy(M)$, for any menu $M$.
\end{lemma}
\proof{Proof.}
To establish the desired claim, it suffices to show that for any supplier $j \in \S$ and for any realization $C_j$ of
$C^M_j = \{ i \in \C : I^M_{ij} = 1 \}$, we have $\expar{ \fproxy_j(M) | C_j^M = C_j} \geq \fcustomized_j(C_j)$. For this purpose, recalling that the customized reward of supplier $j$ can be written as $\fcustomized_j(C_j) = \max_{T_j \subseteq C_j} \{ \sum_{i \in T_j} r_{ij} \cdot \frac{w_{ij}}{1 + \sum_{\ell \in T_j} w_{\ell j}} \}$, suppose that $T_j^*$ is an optimal assortment in this context. Then, $\fcustomized_j(C_j)$ coincides with the standard MNL-based expected revenue of the assortment $T_j^*$, where each ``edge product'' $(i,j)$ is associated with a selling price $r_{ij}$ and a preference weight $w_{ij}$. Letting $R$ be our random revenue in this context, it follows that
\[ \fcustomized_j(C_j) ~~=~~ \int_0^{ \infty } \pr{ R \geq \rho } \mathrm{d} \rho ~~=~~ \int_0^{ \infty } \frac{ \sum_{i \in T_j^* : r_{ij} \geq \rho} w_{ij}}{1 + \sum_{i \in T_j^*} w_{ij}} \mathrm{d} \rho \ . \]
Consequently, it remains to argue that $\prpar{ \fproxy_j(M) \geq \rho | C_j^M = C_j} \geq \frac{ \sum_{i \in T_j^* : r_{ij} \geq \rho} w_{ij}}{1 + \sum_{i \in T_j^*} w_{i j}}$ for every $\rho \geq 0$. Indeed, when $C_j^M = C_j$, a sufficient condition for obtaining a proxy reward of at least $\rho$ is that one or more of the customers in $\{ i \in T_j^* : r_{ij} \geq \rho \}$ qualify. Therefore,
\begin{eqnarray*}
\pr{ \left. \fproxy_j(M) \geq \rho \right| C_j^M = C_j} & \geq & 1 - \pr{ \bigwedge_{i \in T_j^* : r_{ij} \geq \rho} \{ J_{ij} = 0 \} } \\
& = & 1 - \prod_{i \in T_j^* : r_{ij} \geq \rho} \frac{ 1 }{ 1 + w_{ij} } \\
& \geq & 1 - \frac{ 1 }{ 1 + \sum_{i \in T_j^* : r_{ij} \geq \rho} w_{ij} } \\
& \geq & \frac{ \sum_{i \in T_j^* : r_{ij} \geq \rho} w_{ij}}{1 + \sum_{i \in T_j^*} w_{ij}} \ ,
\end{eqnarray*}
where the first equality holds since $\{ J_{ij} \}_{i \in \C}$ are independent, with $J_{ij} \sim \mybern(\hat{w}_{ij})$ and $\hat{w}_{ij} = \frac{ w_{ij} }{ 1 + w_{ij} }$.  
\endproof

\paragraph{Useful representation of \; $\expar{ \fproxy_j(M) }$.} For purposes of subsequent analysis, we derive a closed-form expression for the expected proxy reward $\expar{ \fproxy_j(M) }$ of a single supplier. Specifically, let $\rho_j^{(1)} >  \cdots > \rho_j^{(L_j)} > 0$ be the sequence of distinct positive rewards among $(r_{ij})_{i \in \C}$, and let $\Delta_j^{(\ell)} = \rho_j^{(\ell)} - \rho_j^{(\ell+1)}$ denote the difference between the $\ell$-th and $(\ell+1)$-th elements in this sequence, with the convention that $\rho_j^{(L_j+1)} = 0$. Then, based on arguments similar to those employed while proving Lemma~\ref{lem:proxy_UB}, we know that
\begin{eqnarray}
\ex{ \fproxy_j(M) } &= &\int_0^{ \infty } \pr{ \fproxy_j(M) \geq \rho } \mathrm{d} \rho \nonumber \\
& = & \sum_{\ell \in [L_j]} \Delta_{j}^{(\ell)} \cdot \left( 1 -  \pr{ \bigwedge_{i \in \C : r_{ij} \geq \rho_j^{(\ell)} } \left\{ J_{ij} I^M_{ij} = 0 \right\} } \right) \nonumber \\
& = & \sum_{\ell \in [L_j]} \Delta_{j}^{(\ell)} \cdot \left( 1 - \prod_{i \in \C : r_{ij} \geq \rho_j^{(\ell)} } \left( 1 - \hat{w}_{ij} \cdot \pi_i(j,M_i) \right) \right) \ . \label{eqn:rep_proxy_rew}
\end{eqnarray}

\subsection{The proxy model: A submodular maximization view}
\label{subsec:proxy_submodular}

We proceed by arguing that, with an appropriate set system in place, the computational task of identifying a menu $M$ whose expected proxy reward $\revproxy(M)$ is maximized can be viewed as a submodular maximization problem subject to a single matroid constraint.

\paragraph{Elements and partition matroid.} To define our ground set of elements $\E$, for each customer $i \in \C$ and for every possible menu $M_i \subseteq \S$, we create a corresponding element $(i,M_i)$. It is worth pointing out that this definition leads to $|\C| \cdot 2^{ |\S| }$ elements overall, meaning that $\E$ is implicitly represented. Nevertheless, as explained later on, our algorithmic approach still admits a polynomial-time implementation. To ensure that each customer is offered a single menu, we introduce a partition matroid ${\cal M}$, where out of each set $\E_i = \{ (i,M_i) : M_i \subseteq \S \}$, at most one element can be selected.

\paragraph{Set function.} We now turn our attention to defining a set function $F : 2^{\E} \to \bbR_+$ that identifies with the expected proxy reward $\revproxy$ via a natural bijection between bases of ${\cal M}$ and customer menus. Toward this objective, motivated by representation~\eqref{eqn:rep_proxy_rew} of single-supplier rewards, for every element $(i,M_i)$, supplier $j$, and reward level $\ell$, we define the auxiliary parameter
\[ \beta_{(i,M_i),j,\ell} ~~=~~
\indicatorx \{r_{ij} \geq \rho_j^{(\ell)}\} \cdot \hat{w}_{ij} \cdot \pi_i(j,M_i) \ . \]
With this notation, for an arbitrary collection of elements $\X \subseteq \E$, our set function is given by
\[ F(\X) ~~=~~ \sum_{j \in \S} \sum_{\ell \in [L_j]} \Delta_{j}^{(\ell)} \cdot \left( 1 - \prod_{(i,M_i) \in \X} \left( 1 - \beta_{(i,M_i),j,\ell} \right) \right) \ . \]
To better understand the exact type of equivalence between $F$ and $\revproxy$, note that every base $\X$ of the matroid ${\cal M}$ contains exactly one choice $(i,M_i)$ out of each set $\E_i$. Letting $M^{\X} = (M_i)_{i \in \C}$ denote the corresponding menu, it is easy to verify that $\revproxy( M^{\X} ) = F( \X )$. In the opposite direction, any menu $M$ can be similarly translated to a base $\X^M$ such that $F( \X^M ) = \revproxy( M )$.

\paragraph{Properties of $F$.} The next lemma shows that our reformulation exhibits the monotonicity and submodularity required to employ the greedy algorithm. Generally speaking, these properties follow from the product representation of $F$, in which adding an element makes a nonnegative marginal contribution that decreases as the selected set grows.

\begin{lemma} \label{lem:comb-submodular}
The set function $F : 2^{\E} \to \bbR_+$ is normalized, monotone, and
submodular.
\end{lemma}
\proof{Proof.}
We first observe that, since $F$ is a non-negative combination of the functions $\{ F_{j \ell} \}_{j \in \S, \ell \in [L_j] }$, where $F_{j \ell }(\X) = 1 - \prod_{(i,M_i) \in \X} ( 1 - \beta_{(i,M_i),j,\ell} )$, it suffices to show that each such function is normalized, monotone, and
submodular. Normalization is immediate, and to argue about additional properties, we note that the marginal contribution of every element $(k,M_k) \notin \X$ is
\begin{equation} \label{eqn:F_marginal_cont}
F_{j \ell }(\X \cup \{(k,M_k)\} ) - F_{j \ell }( \X ) ~~=~~ \beta_{(k,M_k),j,\ell} \cdot \prod_{(i,M_i) \in \X} ( 1 - \beta_{(i,M_i),j,\ell} ) \ .
\end{equation}
This quantity is nonnegative, proving monotonicity. In addition, $\prod_{(i,M_i) \in \X} ( 1 - \beta_{(i,M_i),j,\ell} ) \geq \prod_{(i,M_i) \in \hat{\X}} ( 1 - \beta_{(i,M_i),j,\ell} )$ whenever $\X \subseteq \hat{\X}$, implying that the marginal contribution of any element can only decrease as the base set grows, thereby establishing submodularity.  
\endproof

\paragraph{Approximating the proxy model.} Based on the preceding discussion, to obtain a $\frac{ 1 }{ 2 }$-approximation for the proxy model, it remains to derive this guarantee for the submodular maximization problem we have just created. For this purpose, we employ the simple greedy algorithm \citep{NemhauserWF78, CalinescuCPV11}, which is well known to attain an approximation ratio of $\frac{ 1 }{ 2 }$ in the presence of a single matroid constraint. Specialized to our setting, this approach constructs a sequence $\X_0, \X_1, \ldots, \X_{|\C|}$ of independent sets in ${\cal M}$, starting with $\X_0 = \emptyset$. Then, at each iteration $1 \leq k \leq |\C|$, we greedily augment $\X_k = \X_{k-1} \cup \{ (k,M_k) \}$, where the element $(k,M_k)$ is one whose  marginal contribution $F(\X_{k-1} \cup \{(k,M_k)\} ) - F( \X_{k-1} )$ is maximized across all elements whose addition to $\X_{k-1}$ still leaves us with an independent set.

As previously mentioned, a fundamental issue that prevents an immediate application of this idea is that our ground set $\E$ consists of $|\C| \cdot 2^{ |\S| }$ elements, meaning that it may be unclear how to implement the greedy choice $(k,M_k)$ in polynomial time. To resolve this issue, we note that since the partition matroid ${\cal M}$ allows picking at most one element out of each $\E_i = \{ (i,M_i) : M_i \subseteq \S \}$, sets we have already selected from in constructing $\X_{k-1}$ are no longer relevant for choosing $(k,M_k)$, whereas all other sets are relevant for this choice. Consequently, the remaining question is: Given a customer $k \in \C$ for which $\E_k \cap \X_{k-1} = \emptyset$, how do we compute a menu $M_k \subseteq \S$ that maximizes $F(\X_{k-1} \cup \{(k,M_k)\} ) - F( \X_{k-1} )$ over all $2^{ |\S| }$ possible menus? Surprisingly, by plugging representation~\eqref{eqn:F_marginal_cont} for the marginal contributions of elements outside of a given set back into the combined function $F$, we have
\begin{align*}
F(\X_{k-1} \cup & \{(k,M_k)\} ) - F( \X_{k-1} ) \\
& = \sum_{j \in \S} \sum_{\ell \in [L_j]} \Delta_{j}^{(\ell)} \cdot \beta_{(k,M_k),j,\ell} \cdot \prod_{(i,M_i) \in \X_{k-1}} ( 1 - \beta_{(i,M_i),j,\ell} ) \\
& = \sum_{j \in \S} \sum_{\ell \in [L_j]} \Delta_{j}^{(\ell)} \cdot \indicatorx \{r_{kj} \geq \rho_j^{(\ell)}\} \cdot \hat{w}_{kj} \cdot \pi_k(j,M_k) \cdot \prod_{(i,M_i) \in \X_{k-1}} ( 1 - \beta_{(i,M_i),j,\ell} ) \\
& = \sum_{j \in \S} \pi_k(j,M_k) \cdot \underbrace{ \hat{w}_{kj} \cdot \sum_{\ell \in [L_j] : r_{kj} \geq \rho_j^{(\ell)} } \Delta_{j}^{(\ell)} \cdot  \prod_{(i,M_i) \in \X_{k-1}} ( 1 - \beta_{(i,M_i),j,\ell} ) }_{ v_{kj} } \ .
\end{align*}
The important observation is that the cumbersome inner expression, denoted by $v_{kj}$, is independent of the menu $M_k$. Therefore, the menu we are looking for is one that maximizes $\sum_{j \in \S} \pi_k(j,M_k) \cdot v_{kj}$. We have just arrived at a standard MNL-based assortment optimization problem, where preferences are governed by the individual MNL choice model of customer $k$ and selling prices are given by $(v_{kj})_{j \in \S}$. As mentioned in Section~\ref{sec:related-work}, this problem is known to be polynomial-time solvable via revenue-ordered assortments \citep{Talluri04}, for example.

\subsection{Converting proxy reward to inclusive reward}
\label{subsec:comb-conversion}

\paragraph{Intent.} In what follows, we make use of $M^{\mycust*}$, $M^{\myinc*}$, and $M^{\myproxy*}$ to denote optimal menus for the customized, inclusive, and proxy models, respectively. To briefly summarize our main findings up until now, by Lemma~\ref{lem:proxy_UB}, we know that the expected proxy reward $\revproxy(M)$ of any menu $M$ dominates its expected customized reward $\revcustomized( M )$, implying that $\revproxy(M^{\myproxy*}) \geq \revcustomized( M^{\mycust*} )$. Concurrently, as explained in Section~\ref{subsec:proxy_submodular}, by going through an equivalent submodular maximization formulation, the greedy algorithm allows us to efficiently compute an approximate menu $\tilde{M}$ with an expected proxy reward of $\revproxy( \tilde{M} ) \geq \frac{ 1 }{ 2 } \cdot \revproxy(M^{\myproxy*})$. Therefore, to conclude the proof of Theorem~\ref{thm:combinatorial}, it remains to show that $\tilde{M}$ can be efficiently converted into a menu $\hat{M}$ for our original inclusive model, with an expected reward of $\revinclusive( \hat{M} ) \geq \frac{ 1 }{ 2 } \cdot \revproxy( \tilde{M} )$. Indeed, putting all inequalities together,
\[ \revinclusive( \hat{M} ) ~~\geq~~ \frac{ 1 }{ 2 } \cdot \revproxy( \tilde{M} ) ~~\geq~~ \frac{ 1 }{ 4 } \cdot \revproxy(M^{\myproxy*}) ~~\geq~~ \frac{ 1 }{ 4 } \cdot \revcustomized( M^{\mycust*} ) ~~\geq~~ \frac{ 1 }{ 4 } \cdot \revinclusive( M^{\myinc*} ) \ , \]
{where the last inequality follows because the expected reward under the customized model is at least the expected inclusive reward.}
\paragraph{Truncating $\tilde{M}$ to $\hat{M}$.} To create the desired inclusive menu $\hat{M}$, let us focus on a single supplier $j \in \S$, who will be associated with an auxiliary MNL-based assortment optimization problem. Here, our underlying collection of products corresponds to the set of customers $\C$, where in terms of parameters, each customer $i \in \C$ is coupled with a selling price $r_{ij}$ and a preference weight $p_{ij} = \hat{w}_{ij} \cdot \pi_i(j, \tilde{M}_i)$. As such, the optimal expected revenue in this context can be written as
\begin{equation} \label{eqn:aux_comb_Pj} \tag{P$_j$}
\tilde{\rho}_j ~~=~~ \max_{T_j \subseteq \C} \left\{ \sum_{i \in T_j} r_{ij} \cdot \frac{p_{ij}}{1 + \sum_{\ell \in T_j} p_{\ell j}} \right\} \ .
\end{equation}
Our analysis will exploit two well-known properties of optimal assortments,  stating that $\tilde{\rho}_j$ is the unique solution of the equation $\rho_j = \sum_{i \in \C} p_{ij} \cdot [r_{ij} - \rho_j]^+$ and that $\tilde{T}_j = \{ i \in \C : r_{ij} > \tilde{\rho}_j \}$ is an optimal assortment (see, e.g., \citet[Thm.~5.1 and its proof, pp.~133 and~156]{GallegoT19}). We are now ready to convert the greedily produced menu $\tilde{M}$ into an inclusive menu $\hat{M}$, by filtering out edges connecting each supplier $j \in \S$ to customers outside of $\tilde{T}_j$. In other words, the menu displayed to each customer $i \in \C$ is given by $\hat{M}_i = \{ j \in \tilde{M}_i : i \in \tilde{T}_j \}$.

\paragraph{Relating $\revinclusive( \hat{M} )$ and $\revproxy( \tilde{M} )$.} In Lemmas~\ref{lem:cust_vs_tilderho} and~\ref{lem:proxy_vs_tilderho} below, we relate the expected inclusive reward $\expar{ \finclusive_j( C_j^{ \hat{M} } ) }$ of every supplier $j \in \S$ and her expected proxy reward $\expar{ \fproxy_j( \tilde{M} ) }$ by squeezing $\tilde{\rho}_j$ in between, showing that $\expar{ \finclusive_j( C_j^{ \hat{M} } ) } \geq \tilde{\rho}_j \geq \frac{ 1 }{ 2 } \cdot \expar{ \fproxy_j( \tilde{M} ) }$. As such, by summing this inequality across all suppliers, we have $\revinclusive( \hat{M} ) \geq \frac{ 1 }{ 2 } \cdot \revproxy( \tilde{M} )$.

\begin{lemma} \label{lem:cust_vs_tilderho}
$\expar{ \finclusive_j( C_j^{ \hat{M} } ) } \geq \tilde{\rho}_j$, for every supplier $j \in \S$.
\end{lemma}
\proof{Proof.}
We begin by observing that
\begin{equation} \label{eqn:cust_vs_tilderho_eq1}
\ex{ \finclusive_j( C_j^{ \hat{M} } ) } ~~=~~ \ex{ \sum_{i \in C_j^{ \hat{M} } } r_{ij} \cdot \frac{w_{ij}}{1 + \sum_{\ell \in C_j^{ \hat{M} }} w_{\ell j}} } ~~=~~ \sum_{i \in \tilde{T}_j } r_{ij} \cdot \ex{ \frac{w_{ij} I^{ \hat{M} }_{ij}}{1 + \sum_{\ell \in \tilde{T}_j} w_{\ell j} I^{ \hat{M} }_{\ell j}} } \ ,
\end{equation}
where the last equality holds since $C^{ \hat{M} }_j = \{ i \in \tilde{T}_j : I^{ \hat{M} }_{ij} = 1 \}$, by definition of $\hat{M}$. Now, for ease of notation, for every customer $i \in \tilde{T}_j$, let $\mu_{ij} = \expar{ \frac{w_{ij} I^{ \hat{M} }_{ij}}{1 + \sum_{\ell \in \tilde{T}_j} w_{\ell j} I^{ \hat{M} }_{\ell j}} }$ be the inner expectation above, and let $\mu_{0j} = 1 - \sum_{i \in \tilde{T}_j} \mu_{ij}$. Our next step consists of relating $\mu_{ij}$, $\mu_{0j}$, and the preference weight $p_{ij}$ of customer $i$ in formulation~\eqref{eqn:aux_comb_Pj}. The specifics of this relation are given in the next claim, whose proof is presented in Appendix~\ref{app:proof_clm_muij_vs_pij}.

\begin{claim} \label{clm:muij_vs_pij}
$\mu_{ij} \geq p_{ij} \mu_{0j}$, for every $i \in \tilde{T}_j$.
\end{claim}

Proceeding from inequality~\eqref{eqn:cust_vs_tilderho_eq1}, we observe that
\begin{eqnarray}
\ex{ \finclusive_j( C_j^{ \hat{M} } ) } & \geq & \sum_{i \in \tilde{T}_j } r_{ij} \mu_{ij} \nonumber \\
& = & \tilde{\rho}_j \cdot \sum_{i \in \tilde{T}_j } \mu_{ij} + \sum_{i \in \tilde{T}_j } [r_{ij} - \tilde{\rho}_j]^+ \cdot \mu_{ij} \label{eqn:cust_vs_tilderho_eq2} \\
& \geq & \tilde{\rho}_j \cdot \sum_{i \in \tilde{T}_j } \mu_{ij} + \mu_{0j} \cdot \sum_{i \in \tilde{T}_j } [r_{ij} - \tilde{\rho}_j]^+ \cdot p_{ij} \label{eqn:cust_vs_tilderho_eq3} \\
& = & \left( \sum_{i \in \tilde{T}_j } \mu_{ij} + \mu_{0j} \right) \cdot \tilde{\rho}_j \label{eqn:cust_vs_tilderho_eq4} \\
& = & \tilde{\rho}_j \ .
\end{eqnarray}
Here, equality~\eqref{eqn:cust_vs_tilderho_eq2} holds since $\tilde{T}_j = \{ i \in \C : r_{ij} > \tilde{\rho}_j \}$. Inequality~\eqref{eqn:cust_vs_tilderho_eq3} is obtained by plugging in Claim~\ref{clm:muij_vs_pij}. Equality~\eqref{eqn:cust_vs_tilderho_eq4} follows by recalling that $\tilde{\rho}_j = \sum_{i \in \C} p_{ij} \cdot [r_{ij} - \tilde{\rho}_j]^+$ and noting that $[r_{ij} - \tilde{\rho}_j]^+ = 0$ for every $i \notin \tilde{T}_j$. 
\endproof

\begin{lemma} \label{lem:proxy_vs_tilderho}
$\expar{ \fproxy_j( \tilde{M} ) } \leq 2 \tilde{\rho}_j$, for every supplier $j \in \S$.
\end{lemma}
\proof{Proof.}
By circling back to definition~\eqref{eqn:f_proxy_definition}, we recall that $\fproxy_j( \tilde{M} ) = \max_{i \in \C} \{ r_{ij} I_{ij}^{ \tilde{M} } J_{ij} \}$, where $\{ J_{ij} \}_{i \in \C}$ are independent indicators with $J_{ij} \sim \mybern(\hat{w}_{ij})$, independent of the customer choices $\{ I_{ij}^{ \tilde{M} } \}_{i \in \C}$. We first observe that
\[ \fproxy_j( \tilde{M} ) ~~\leq~~ \tilde{\rho}_j + \max_{i \in \C} \left\{ [r_{ij} - \tilde{\rho}_j]^+ \cdot I_{ij}^{ \tilde{M} } J_{ij} \right\} ~~\leq~~ \tilde{\rho}_j + \sum_{i \in \tilde{T}_j} [r_{ij} - \tilde{\rho}_j]^+ \cdot I_{ij}^{ \tilde{M} } J_{ij} \ , \]
where the last inequality holds since $\{ [r_{ij} - \tilde{\rho}_j]^+ \cdot I_{ij}^{ \tilde{M} } J_{ij} \}_{i \in \C}$ are non-negative random variables, and since $\tilde{T}_j = \{ i \in \C : r_{ij} > \tilde{\rho}_j \}$. By taking expectations,
\begin{eqnarray}
\ex{ \fproxy_j( \tilde{M} ) }  & \leq & \tilde{\rho}_j + \sum_{i \in \tilde{T}_j} [r_{ij} - \tilde{\rho}_j]^+ \cdot \ex{ I_{ij}^{ \tilde{M} } J_{ij} } \nonumber \\
& = & \tilde{\rho}_j + \sum_{i \in \tilde{T}_j} [r_{ij} - \tilde{\rho}_j]^+ \cdot \ex{ I_{ij}^{ \tilde{M} } } \cdot \ex{ J_{ij} } \label{eqn:proxy_vs_tilderho_eq1} \\
& = & \tilde{\rho}_j + \sum_{i \in \tilde{T}_j} p_{ij} \cdot [r_{ij} - \tilde{\rho}_j]^+ \label{eqn:proxy_vs_tilderho_eq2} \\
& = & 2 \tilde{\rho}_j \ . \label{eqn:proxy_vs_tilderho_eq3}
\end{eqnarray}
Here, equality~\eqref{eqn:proxy_vs_tilderho_eq1} holds since $I_{ij}^{ \tilde{M}}$ and $J_{ij}$ are independent. Equality~\eqref{eqn:proxy_vs_tilderho_eq2} follows by noting that $\expar{ I_{ij}^{ \tilde{M} } } \cdot \expar{ J_{ij} } = \pi_i(j, \tilde{M}_i) \cdot \hat{w}_{ij} = p_{ij}$. Finally, equality~\eqref{eqn:proxy_vs_tilderho_eq3} is obtained by recalling $\tilde{\rho}_j = \sum_{i \in \C} p_{ij} \cdot [r_{ij} - \tilde{\rho}_j]^+$ and noting that $[r_{ij} - \tilde{\rho}_j]^+ = 0$ for every $i \notin \tilde{T}_j$. 
\endproof

\section{Linear Programming-based Approach}\label{sec:lpbased_approach}
In what follows, we establish Theorem~\ref{thm:main_algorithm} by developing an LP-based polynomial-time algorithm that approximates the inclusive model within factor $\frac{ 1 }{ 2 }$. Toward this objective, in Section~\ref{sec:preliminaries}, we introduce a number of technical ideas from the assortment optimization literature that will allow us to derive our linear programming relaxation and, consequently, to analyze it. In Section~\ref{sec:customized}, we start by presenting our main relaxation that upper bounds both the optimal customized and inclusive values. Later, in Section~\ref{subsec:customized_thm}, we utilize supplier-specific reward thresholds to obtain a fractional solution whose inclusive reward value is at least half of the LP optimal value, thereby concluding our analysis. Finally, in Section~\ref{subsec:related_results}, we derive a guarantee for the customized model and we establish the exact {reward gap} between both models.

\subsection{Preliminaries}\label{sec:preliminaries}

The main purpose of this section is to lay down a number of technical ideas that will be important in describing our algorithmic approach and in analyzing its performance guarantees. In Section~\ref{subsec:the_MNL_choice_polyhedron}, we introduce the so-called MNL-choice polyhedron, and discuss folklore ideas for translating any point in this polyhedron to randomized assortments, preserving the choice probabilities of all alternatives. It is important to mention that this randomized approach and its analysis are not new to the assortment optimization literature, and should mostly be attributed to the work of~\citet[Thm.~1]{Topaloglu13} and~\citet[Sec~5.1]{GallegoRS15}.
Then, in Section~\ref{subsec:equivalent_formulation}, we present an equivalent formulation of our models of interest, offering a workable representation of the two-step process that creates random customer-supplier matchings.

\subsubsection{The MNL-choice polyhedron and randomized assortments.}\label{subsec:the_MNL_choice_polyhedron}

To flesh out some of the main ideas, let us consider a generic instantiation of the MNL choice model, where we are given a ground set of alternatives $[n]=\{1,\ldots,n\}$, whose preference weights are respectively denoted by $u_1, \ldots, u_n$, whereas that of the outside option is $u_0 = 1$. According to the MNL model, with respect to any assortment $A \subseteq [n]$, the choice probability of any alternative $j \in [n]$ is given by $\pi(j,A) = \frac{ u_j }{ 1 + u(A) } \cdot \indicatorx[j \in A]$, with the convention that $u(A) = \sum_{k\in A}u_k$. 
By temporarily denoting $x_j = \pi(j,A)$, one can easily verify that $\frac{ x_j }{ u_j } \leq 1 - \sum_{\ell \in [n]} x_\ell$ for every alternative $j \in [n]$.
Now, let us examine the polyhedron induced by the conjunction of these inequalities, namely,
\begin{equation*}
    P_{\mathrm{MNL}} ~~=~~ \left\{ x \in \bbR^n_+ : \ \frac{ x_j }{ u_j } \leq 1 - \sum_{\ell \in [n]} x_\ell,  \ \text{for all} \ j \in [n] \right\} \ .
\end{equation*}
As explained above, these constraints are satisfied by the choice probabilities of any assortment. However, suppose we ask the following question in the opposite direction: Given a point $x \in P_{\mathrm{MNL}}$, is there an assortment $A(x)$ where the choice probabilities of the various alternatives are precisely $x_1, \ldots, x_n$? 
Formally, we want to efficiently construct a distribution $\mathcal{D}(x)$ over $n + 1$ assortments such that, upon sampling from $\mathcal{D}(x)$, the probability of choosing each alternative $j \in [n]$ is precisely $x_j$. We begin by elaborating on the specifics of this construction.

\paragraph{The distribution ${\mathcal{D}(x)}$.}
Without loss of generality, we assume that alternatives are indexed such that
\begin{equation} \label{eqn:order_ratios}
    \frac{x_1}{u_1} ~~\geq~~ \frac{x_2}{u_2} ~~\geq~~ \cdots ~~\geq~~ \frac{x_n}{u_n} \ ,
\end{equation}
with the convention that $x_0 = 1 - \sum_{j \in [n]} x_j$ and $u_0 = 1$. 
As such, our sample space will be comprised of the assortments $A_0, \ldots, A_n$, where each $A_j$ is the prefix of the sequence $1, \ldots, n$, consisting of its first $j$ alternatives, i.e., $A_0 = \emptyset$ and $A_j = [j]$ for $j\geq1$. The respective probabilities of picking each of these assortments will be denoted for convenience as $\psi_0, \ldots, \psi_n$, given by:
\begin{itemize}
    \item $\psi_j = (\frac{x_j}{u_j} - \frac{x_{j+1}}{u_{j+1}}) \cdot \sum_{\ell = 0}^j u_\ell$, for $j = 0, \ldots, n - 1$.

    \item $\psi_n = \frac{x_n}{u_n} \cdot \sum_{\ell = 0}^n u_\ell$.
\end{itemize}

\paragraph{Analysis.} Let us first observe that the distribution $\mathcal{D}(x)$ is well-defined. Indeed, $\psi_j \geq 0$ for every $j \in [n]$, since according to the ordering~\eqref{eqn:order_ratios}, we have $\frac{x_j}{u_j} \geq \frac{x_{j+1}}{u_{j+1}}$. In addition,
\[ \psi_0 ~~=~~ \left( \frac{x_0}{u_0} - \frac{x_1}{u_1} \right) \cdot u_0 ~~=~~ 1 - \sum_{\ell \in [n]} x_\ell - \frac{x_1}{u_1} ~~\geq~~ 0 \ , \]
where the last inequality holds since $x \in P_{\mathrm{MNL}}$, implying in particular that $\frac{ x_1 }{ u_1 } \leq 1 - \sum_{\ell \in [n]} x_\ell$. The following lemma, whose proof is given in Appendix~\ref{app:proof_assortment_distribution_is_well_defined}, shows that $\psi_0, \ldots, \psi_n$ can be viewed as probabilities, as these parameters sum to $1$.

\begin{lemma}\label{lem:assortment_distribution_is_well_defined}
    $\sum_{j = 0}^n \psi_j = 1$.
\end{lemma}

Now, consider the scenario in which we randomly sample an assortment from the distribution $\mathcal{D}(x)$ and select an alternative out of this  assortment according to the MNL model.
We argue that each alternative $j \in [n]$ is selected with probability $x_j$. 
To formalize this claim, we present the following lemma, which confirms that $x_j$ indeed identifies with the choice probability $\exparsub{A \sim \mathcal{D}(x)}{\pi (j, A)}$, recalling that $\pi(j, A) = \frac{ u_j }{ 1 + u(A) } \cdot\indicatorx[j \in A]$ denotes the probability of selecting alternative $j$ out of the assortment $A$.
This result, whose proof is provided in Appendix~\ref{app:proof_assortment_distribution_realizes_LP_solution}, will be crucial for understanding how such randomized assortments translate into choice probabilities.

\begin{lemma}\label{lem:assortment_distribution_realizes_LP_solution}
    $\exparsub{A \sim \mathcal{D}(x)}{\pi (j, A)} = x_j$, for every alternative $j \in [n]$.
\end{lemma}

\subsubsection{Reformulation: From menus to choice probabilities.} \label{subsec:equivalent_formulation}

Circling back to Section~\ref{subsec:description_of_the_problem}, recall that whether we consider the customized model or the inclusive model, our objective is to determine a menu $M = (M_i)_{i \in \C}$ whose expected reward $\rev(M)$ is maximized.
In what follows, we reformulate both models in terms of an extended MNL-choice polyhedron, making these settings continuous in nature.
This representation will be instrumental in expanding the toolkit we have at our disposal, especially for the use of linear programming relaxations.

\paragraph{The customers' polyhedron ${P^\C}$.}
Consider a customer $i \in \C$, and recall that her individual MNL choice polyhedron is given by
\begin{equation} \label{def:P^C_i}
    P^\C_i ~~=~~ \left\{ x \in \bbR^{|\S|}_+ :\, \frac{ x_j }{ u_{ij} } \leq 1 - \sum_{k \in \S} x_k, \, \, \text{for all } j \in \S \right\} \ .
 \end{equation}
Now, our polyhedron of interest $P^\C$ jointly describes the choice probabilities of all customers, specified by the Cartesian product of $P^\C_i$ over all customers $i \in \C$, namely 
\begin{equation} \label{eqn:polyhedron_P_C}
    P^\C ~~=~~ \left\{ x \in {\bbR}_+^{\vert \C \vert \times \vert \S \vert} :\, x_{i \cdot} \in P^\C_i \, , \, \, \text{for all } i \in \C \right\} \ .
\end{equation}
\paragraph{The suppliers' polyhedron ${P^\S}$.} Even though the next polyhedron is not used in the current section, its definition will be important in later sections. Similarly to the customers' polyhedron, for each supplier $j \in \S$, we define
\[ P^\S_j ~~=~~ \left\{ x \in \bbR^{|\C|}_+ :\, \frac{ x_i }{ w_{ij} } \leq 1 - \sum_{\ell \in \C} x_\ell, \, \, \text{for all } i \in \C \right\} \ , \]
with the combined polyhedron $P^\S$ given by
\begin{equation*} \label{eqn:polyhedron_P_S}
    P^\S ~~=~~ \left\{ x \in {\bbR}_+^{\vert \C \vert \times \vert \S \vert} : x_{\cdot j} \in P^\S_j \, , \, \text{for all } j \in \S \right\} \ .
\end{equation*}
As a side note, a particularly useful and easy-to-verify property of the polyhedra $P^\C$ and $P^\S$ is that they are both downward-closed.
That is, when $x \in P^\C$ and $0 \leq y \leq x$ coordinate-wise, then $y \in P^\C$ as well, and the same goes for $P^\S$.
This property will come in handy in subsequent sections, mostly to imply that we are preserving feasibility along various structural manipulations.

\paragraph{Analog of expected reward.} We proceed by proposing a convenient way to express the notion of ``expected reward'' as a function of points in the customers' polyhedron $P^\C$ rather than in terms of menus. To this end, given a point $x \in P^\C$, we will define the function $\rev_P(x)$ to simultaneously capture the reward gained across all suppliers, whether one considers the customized or the inclusive model.
With this objective in mind, let $I^x_{ij}$ be an indicator random variable with $I^x_{ij} \sim \mybern(x_{ij})$, such that for every supplier $j \in \S$, the indicators $(I^x_{ij})_{i \in \C}$ are mutually independent.
Importantly, we make no assumptions whatsoever about the joint distribution of $(I^x_{ij})_{j \in \S}$ for any customer $i \in \C$, since this information is not required for our analysis. Now, for any supplier $j \in \S$, consider the random set of customers for whom the indicator $I^x_{ij}$ equals 1, i.e., $C^x_j = \{ i \in \C : I^x_{ij} = 1 \}$.
With these ingredients, we introduce the function $\rev_P: P^{\C} \to \mathbb{R}$,   {a continuous analogue of \eqref{eqn:definition_reward},} defined by
\begin{equation}\label{definition:R^P}
    \rev_P(x) ~~=~~ \sum_{j \in S} \ex{f_j(C^x_j)}\;,
\end{equation}
where the expectation is taken over the randomness in $(I^x_{ij})_{i \in \C, j \in \S}$. {When necessary, we distinguish both cases by writing $\revcustP$ for $f_j=\fcustomized_j$ and $\revincP$ for $f_j=\finclusive_j$.} 

\paragraph{Problem reformulation.}
In what follows, we argue that $\rev_P$ can play the role of an alternative function to capture our expected reward. Specifically, recalling that in terms of menus,
the original problem we are addressing is given by
\begin{equation} \label{reformulation1} \tag{P}
\begin{array}{llll}
\max        & {\displaystyle \rev(M) } & \\
\text{s.t.} & {\displaystyle M = (M_1, \ldots, M_{|\C|}) } & \\
& {\displaystyle  M_i \subseteq \S} & \forall \, i \in \C
\end{array}
\end{equation}
we replace its objective $\rev$ by $\rev_P$, focusing on the polyhedron $P^\C$ as our feasible region, to obtain
\begin{equation} \label{eqn:problem_R} \tag{$\mathrm{R}$}
\begin{array}{lll}
\max        & {\displaystyle \rev_P(x) } \\
\text{s.t.} & {\displaystyle x \in P^\C } 
\end{array}
\end{equation}
{When clear from context, we omit which reward model~\eqref{eqn:problem_R} refers to.} The first claim we establish is that~\eqref{eqn:problem_R} is a relaxation of~\eqref{reformulation1}, as stated in the next lemma.

\begin{lemma}\label{lem:reformulation_relaxation}
    $\opt\eqref{eqn:problem_R} \geq \opt\eqref{reformulation1}$.
\end{lemma}
\proof{Proof.}
Let $M^*$ be an optimal menu for~\eqref{reformulation1}. To prove the desired claim, we will show how to define a corresponding point $x \in P^\C$ such that {$\expar{f_j(C^x_j)} = \expar{f_j(C^{M^*}_j)}$} for every supplier $j \in \S$. Our candidate point $x \in \bbR_+^{|\C| \times |\S|}$ is given by $x_{ij} = \pi_i (j, M^*_i)$ for all pairs $(i,j) \in \C \times \S$.
Namely, each vector $x_{i \cdot}$
represents the choice probabilities according to customer $i$'s MNL model with respect to her menu $M^*_i$.

We initially verify that $x \in P^\C$ by showing that $x_{i \cdot} \in P^\C_i$ for every customer $i \in \C$.
According to definition~\eqref{def:P^C_i}, we should show that $x$ is non-negative, which is obvious, and that for every supplier $j \in \S$, we have $\frac{x_{ij}}{u_{ij}} \leq 1 - \sum_{k \in \S} x_{i k}$.
To prove this relation, we consider two cases:
\begin{itemize}
    \item When $j \in M^*_i$, we have
    \begin{eqnarray*}
        && \frac{x_{ij}}{u_{ij}} ~~=~~ \frac{1}{1 + \sum_{k \in M^*_i} u_{i k}} ~~=~~ \pi_i( 0, M^*_i) \\
        && \qquad \qquad \qquad =~~ 1 - \sum_{k \in M^*_i} \pi_i( k, M^*_i) ~~=~~ 1 - \sum_{k \in M^*_i} x_{i k} ~~=~~ 1 - \sum_{k \in \S} x_{i k} \ ,
    \end{eqnarray*}
    where the last equality holds since $x_{ik} = 0$ for all $k \in \S \setminus M^*_i$.
    
    \item When $j \in \S \setminus M^*_i$, we have 
    \[ \frac{x_{ij}}{u_{ij}} ~~=~~ 0 ~~\leq~~ 1 - \sum_{k \in \S} \pi_i(k, M^*_i) ~~=~~ 1 - \sum_{k \in \S} x_{i k} \ .  \]   
\end{itemize}

Knowing that $x$ is a feasible solution to~\eqref{eqn:problem_R}, 
it remains to show that $\expar{f_j(C^x_j)} = \expar{f_j(C^{M^*}_j)}$ for every supplier $j \in \S$, since this claim implies that  
\[ \opt\eqref{eqn:problem_R} ~~\geq~~ \rev_P(x) ~~=~~ \sum_{j \in \S} \ex{f_j(C^x_j)} ~~=~~ \sum_{j \in \S} \ex{f_j(C^{M^*}_j)} ~~=~~ \rev(M^*) ~~=~~ \opt\eqref{reformulation1} \ . \]
To this end, we argue that $C^x_j$ and $C^{M^*}_j$ are identically distributed subsets of customers. Let us first recall that $C^x_j = \{ i \in \C: I^x_{ij} = 1 \}$ and $C^{M^*}_j = \{ i \in \C: I^{M^*}_{ij} = 1 \}$.
We know that the indicators  $I^x_{ij}$ and $I^{M^*}_{ij}$ are successful with probability $x_{ij} = \pi_i (j, M^*_i)$. In addition, $(I^{x}_{ij})_{i \in \C}$ and $(I^{M^*}_{ij})_{i \in \C}$ are mutually independent, since: (1)~Each $I^{M^*}_{ij}$ indicates whether customer $i$ selects supplier $j$ or not, and customers' decisions are independent of one another; (2)~By definition, $(I^x_{ij})_{i \in \C}$ are mutually independent. Thus, $C^x_j$ and $C^{M^*}_j$ share the same marginal distribution, and therefore, $\expar{f_j(C^x_j)} = \expar{f_j(C^{M^*}_j)}$. 
\endproof

\paragraph{Solution concept: Distribution over menus.}
To better understand the usefulness of relaxation~\eqref{eqn:problem_R}, let us ask the following question: Given any feasible solution $x$ to~\eqref{eqn:problem_R}, can it be ``translated'' into a menu $M^x$ whose expected reward $\rev(M^x)$ is comparable to $\rev_P(x)$?
At least initially, it is unclear how to deterministically compute such a menu, and thus, we address this question by computing a \emph{distribution} over menus, $\mathcal{D}(x)$.
For this purpose, based on the discussion in Section~\ref{subsec:the_MNL_choice_polyhedron}, given a point $x_{i \cdot} \in P^\C_i$ for each customer $i \in \C$, we can efficiently define a distribution $\mathcal{D}(x_{i \cdot})$ of menus such that $\exsubpar{M_i \sim \mathcal{D}(x_{i\cdot})}{\pi_i (j, M_i)} = x_{ij}$ for each supplier $j \in \S$.
Namely, $x_{i \cdot}$ represents the MNL-choice probabilities of customer $i$ with respect to a random menu sampled from $\mathcal{D}(x_{i \cdot})$, in expectation.
Next, $\mathcal{D}(x)$ will be defined as the joint distribution of $\{ \mathcal{D}(x_{i \cdot}) \}_{i \in \C}$, each independently constructed, and the expected reward of a random menu $M$ sampled from $\mathcal{D}(x)$ will be denoted by $\rev(\mathcal{D}(x)) = \exparsub{M \sim \mathcal{D}(x)}{\rev(M)}$.
The upcoming claim, whose proof is provided in Appendix~\ref{app:proof_lem_randomized_assortment}, shows that this measure identifies with $\rev_P(x)$.

\begin{lemma}\label{lem:randomized_assortment}
    $\rev(\mathcal{D}(x)) = \rev_P(x)$.
\end{lemma}

\paragraph{Remark: The opposite direction.}
Even though we will be treating problem~\eqref{eqn:problem_R} as a relaxation of~\eqref{reformulation1}, we mention in passing that it is actually an exact reformulation. To verify this claim, let $x^*$ be an optimal solution to~\eqref{eqn:problem_R}.
Then, 
\begin{equation*}
    \exsub{M \sim \mathcal{D}(x^*)}{\rev(M)} ~~=~~ \rev(\mathcal{D}(x^*)) ~~=~~ \rev_P(x^*) ~~=~~ \opt\eqref{eqn:problem_R} \ ,
\end{equation*}
where the second equality follows from Lemma~\ref{lem:randomized_assortment}. Therefore, there exists a deterministic menu $M^{x^*}$ within the sample space of $\mathcal{D}(x^*)$ for which $\rev(M^{x^*}) \geq \opt\eqref{eqn:problem_R}$, implying that $\opt\eqref{reformulation1} \geq \opt\eqref{eqn:problem_R}$.
Combined with Lemma~\ref{lem:reformulation_relaxation}, where the opposite inequality is established, we arrive at the desired equivalence.

\begin{corollary}\label{cor:reformulation}
    $\opt\eqref{reformulation1} = \opt\eqref{eqn:problem_R}$.
\end{corollary}
{In the remainder, we write $\optcust$ and $\optinc$ to denote the corresponding optimal values for the customized and inclusive models, respectively.}
\subsection{Main Algorithm}\label{sec:customized}

In this section, we devise a constant-factor approximation for the inclusive and customized models, showing how to efficiently compute a random menu whose expected reward is within factor $\frac{1}{2}$ of the optimal customized value, as stated in Theorem~\ref{thm:main_algorithm}.
Towards this objective, in Section~\ref{subsec:customized_relaxation_2}, we first propose an equivalent representation for the objective function of the customized reformulation~\eqref{eqn:problem_R}, albeit relying on exponentially-many decision variables.
Subsequently, we exploit this representation to develop a linear relaxation featuring polynomially-many variables.
Finally, in Section~\ref{subsec:customized_thm}, we prove that an optimal solution to the latter relaxation garners an expected inclusive reward of at least $\frac{1}{2} \cdot \optcust\eqref{eqn:problem_R}$. Crucial to our analysis is the solution of a fractional variant of the fixed point equation~\eqref{eqn:aux_comb_Pj}. As in the combinatorial approach, we bound the gain of the inclusive reward and the linear program relaxation with respect to this fixed point value.

\subsubsection{A linear relaxation with polynomially-many variables.} \label{subsec:customized_relaxation_2}

Our first step toward arriving at the desired linear relaxation is to better understand the objective function $\revcustP$ of formulation~\eqref{eqn:problem_R}. {This is crucial for both reward models since any relaxation for the customized model is also a relaxation for the inclusive model. 
In the customized model, 
\[
\fcustomized_j(C_j) ~~=~~ \max_{T_j \subseteq C_j} \left\{ \sum_{i \in T_j} r_{ij} \cdot \frac{w_{ij}}{1 + \sum_{\ell \in T_j} w_{\ell j}} \right\}
\]
represents the {best possible} expected reward gained from supplier $j$, conditional on $C_j$ being her set of selecting customers.
In other words, $\fcustomized_j(C_j)$ stands for the optimal revenue in the classical MNL-based assortment optimization problem, where each customer $i \in C_j$ can be viewed as a product, associated with a preference weight of $w_{ij}$ and a price of $r_{ij}$. Therefore, based on the well-known LP-formulation of this problem (see, e.g., \citet[Sec.~5.11.1]{GallegoT19}), we can alternatively express $\fcustomized_j(C_j)$ as the optimal value of the following linear program:
\begin{alignat}{3}
\fcustomized_j(C_j) ~~=~~ &  \max &&\quad   \sum_{i \in C_j} r_{ij} y_{ij}  \label{eqn:assortment_optimization} \tag{LP$_{\mathrm{MNL}}$}\\
 &\;\text{s.t.} &&\quad y_{\cdot j} \in P^\S_j \notag\\
&&&\quad y_{ij} = 0 \qquad \forall \,  i \in \C \setminus C_j\,. \label{def:P^{C_j}_j}
\end{alignat}
Now, letting $y^*_{\cdot j}(C_j)$ be an optimal solution to~\eqref{eqn:assortment_optimization}, which clearly depends on $C_j$, we can rewrite the reward function $\revcustP(x)$ as follows: 
\begin{eqnarray}
    \revcustP(x) & = & \sum_{j \in \S} \ex{\fcustomized_j(C^x_j)} \nonumber \\
     & = & \sum_{j \in \S} \sum_{C_j \subseteq \C} \pr{C^x_j = C_j} \cdot \fcustomized_j(C_j) \nonumber \\
    & = & \sum_{j \in \S} \sum_{C_j \subseteq \C} \pr{C^x_j = C_j} \cdot \sum_{i \in C_j} r_{ij} y^*_{ij}(C_j) \nonumber \\
    & = & \sum_{(i, j) \in \E} r_{ij} \cdot \sum_{C_j \subseteq \C} \pr{C^x_j = C_j} \cdot y^*_{ij}(C_j) \ , \label{equality:intermediate}
\end{eqnarray}
where the last equality holds due to constraint~\eqref{def:P^{C_j}_j}, with the simplifying notation $\E = \C\times\S$.

Our next step proceeds by analyzing the objective value $\revcustP(x^*)$ of an optimal solution $x^*$ to formulation~\eqref{eqn:problem_R}. To this end, for every pair $(i, j) \in \E$, let us introduce the auxiliary variable $\hat{y}_{ij} = \sum_{C_j \subseteq \C} \prpar{C^{x^*}_j = C_j} \cdot y^*_{ij}(C_j)$, which is precisely the inner summation in representation~\eqref{equality:intermediate}, meaning that $\revcustP(x^*) = \sum_{(i, j) \in \E} r_{ij} \hat{y}_{ij}$. In the next lemma, whose proof appears in Appendix~\ref{app:auxiliary_properties}, we uncover two important properties of these variables. To avoid cumbersome expressions, we recall the shorthand notation $\hat{w}_{ij} = \frac{w_{ij}}{1+w_{ij}}$.

\begin{lemma} \label{lem:auxiliary_properties}
The vector $\hat{y}$ satisfies the next two properties:
    \begin{enumerate}
        \item $\hat{y} \in P^\S$.
        \item $\hat{y}_{ij} \leq \hat{w}_{ij} x^*_{ij}$, for every $(i, j) \in \E$.
    \end{enumerate}
\end{lemma}

\paragraph{The linear relaxation.}
The above-mentioned properties motivate us to consider the next linear program, obtained by altering formulation~\eqref{eqn:problem_R} such that each inner summation $\sum_{C_j \subseteq \C} \prpar{C^x_j = C_j} \cdot y^{*}_{ij}(C_j)$ is replaced by the decision variable $y_{ij}$. Additionally, we embed the structural properties of Lemma~\ref{lem:auxiliary_properties}, tweaking item~2 to be written as an equality, to obtain: 
\begin{equation} \label{eqn:customized_lp} \tag{$\mathrm{LP}$}
\begin{array}{llll}
\max  & {\displaystyle \sum_{(i, j) \in \E} r_{ij} y_{ij}} \\
\text{s.t.} & {\displaystyle y_{ij} = \hat{w}_{ij} x_{ij}} &\qquad \forall \,  (i, j) \in \E \\
& {\displaystyle y \in P^\S \text{, } x \in P^\C} & 
\end{array}
\end{equation}
The important observation is that~\eqref{eqn:customized_lp} is indeed a relaxation of~\eqref{eqn:problem_R} {for the customized model}. 
To verify this claim, we know  by Lemma~\ref{lem:auxiliary_properties} that $\hat{y} \in P^\S$ and that $\hat{y}_{ij} \leq \hat{w}_{ij} x^*_{ij}$, for every $(i, j) \in \E$. However, since the {customers}' polyhedron $P^\C$ is downward-closed and since $x^* \in P^\C$, it is not difficult to see that the value of each $x_{ij}^*$ can be decremented until this constraint becomes tight. Letting $\hat{x}$ be the resulting vector, we have just ensured that $(\hat{x}, \hat{y})$ constitutes a feasible solution to~\eqref{eqn:customized_lp}. Consequently, $\optcust\eqref{eqn:customized_lp} \geq \sum_{(i, j) \in \E} r_{ij} \hat{y}_{ij} = \optcust\eqref{eqn:problem_R}$, where the last equality combines representation~\eqref{equality:intermediate} and the definition of $\hat{y}_{ij}$. 

\begin{corollary} \label{cor:customized_lp_relaxation}
    {$\opt\eqref{eqn:customized_lp} \geq \opt^{\mathrm{cust}}\eqref{eqn:problem_R}\geq\opt^{\mathrm{inc}}\eqref{eqn:problem_R}$.}
\end{corollary}

\subsubsection{Final algorithm and its analysis.}\label{subsec:customized_thm}

In the remainder of this section, Theorem~\ref{thm:main_algorithm} will be established by understanding how to connect the dots around the resulting relaxation in a useful way. We remind the reader that our goal is to efficiently compute a point $\tilde{x}\in P^\C$ such that
\begin{equation}\label{eqn:main_guarantee}
    \revincP(\tilde{x})\geq \frac{1}{2}\cdot\optcust\eqref{eqn:problem_R}.
\end{equation}
To construct such a point, we make use of our linear programming relaxation to derive an alternative version of \eqref{eqn:aux_comb_Pj}. Formally, letting $(x^*,y^*)$ be any optimal solution of~\eqref{eqn:customized_lp}, we define $p^*_{ij}=\hat{w}_{ij}\cdot x^*_{ij}$ for all $(i,j)\in\E$ and $\rho^*_j$ as  
\begin{equation}\label{eqn:fixed_point}
\rho^*_j=\max_{T_j\subseteq\C}\left\{ \sum_{i \in T_j} r_{ij} \cdot \frac{p^*_{ij}}{1 + \sum_{\ell \in T_j} p^*_{\ell j}} \right\} \ . \tag{$P^*_j$}
\end{equation}
Note that the main difference between \eqref{eqn:fixed_point} and \eqref{eqn:aux_comb_Pj} is that the former considers $p^*_{ij}=\hat{w}_{ij}\cdot x^*_{ij}$ whereas the latter uses $p_{ij} = \hat{w}_{ij}\cdot\pi_i(j,\tilde M)$ for some menu $\tilde M$. As in the combinatorial approach, our analysis will exploit  the well-known property that $\rho^*_j$ is the unique solution of the equation $\rho_j = \sum_{i \in \C} p^*_{ij} \cdot [r_{ij} - \rho_j]^+$  (see, e.g., \citet[Thm.~5.1 and its proof, pp.~133 and~156]{GallegoT19}). 

\paragraph{Algorithm.} Compute an optimal solution $(x^*,y^*)$ of~\eqref{eqn:customized_lp}.  Construct $\tilde x$ by truncating $x^*$ as follows: For every $(i,j)\in \E$, $\tilde{x}_{ij}=x^*_{ij}$ if $r_{ij}>\rho^*_j$, and $\tilde x_{ij}=0$, otherwise.

\paragraph{Analysis.} To prove that $\tilde x$ satisfies inequality \eqref{eqn:main_guarantee}, we first note that $\tilde x$ lies in $P^\C$ because $0\leq\tilde x\leq x^*\in P^\C$ and $P^{\C}$ is downward-closed. Moreover, as the next result shows, the values $\{\rho^*_j\}_{j\in\S}$ serve as a lower bound for the inclusive reward obtained by $\tilde x$ and, up to a factor of 2, as an upper bound the optimal value of \eqref{eqn:customized_lp}.
\begin{lemma}\label{lem:properties_main}
The point $\tilde x$ and values $\{\rho^*_j\}_{j\in\S}$  satisfy:
\begin{enumerate}
    \item $\revincP(\tilde x)\geq \sum_{j\in\S}\rho^*_j$\;.
    \item $\opt\eqref{eqn:customized_lp}\leq 2\cdot \sum_{j\in\S}\rho^*_j$\;.
\end{enumerate}
\end{lemma}
The desired guarantee \eqref{eqn:main_guarantee} follows immediately: 
\[
\revincP(\tilde x)\geq \sum_{j\in\S}\rho^*_j
\geq \frac{1}{2}\cdot \opt\eqref{eqn:customized_lp}
\geq  \frac{1}{2}\cdot \optcust\eqref{eqn:problem_R}, 
\]
where the first and second inequalities hold due to Lemma~\ref{lem:properties_main}, and we use Corollary~\ref{cor:customized_lp_relaxation} to conclude the last inequality.

\begin{proof}{Proof of Lemma~\ref{lem:properties_main}.}
Recall that, for any $x_{ij}\in[0,1]$, we denote $I_{ij}^x \sim \mybern(x_{ij})$. Define $A_j = \{i\in\C:\ r_{ij}>\rho^*_j\}$. 
For the first property we have 
\begin{align}
\ex{\finclusive_j(C^{\tilde x})} = \ex{\sum_{i\in\C}r_{ij}\cdot\frac{w_{ij}I^{\tilde x}_{ij}}{1+\sum_{i\in\C}w_{ij}I^{\tilde x}_{ij}}} = \ex{\sum_{i\in A_j}r_{ij}\cdot\frac{w_{ij}I^{x^*}_{ij}}{1+\sum_{i\in A_j}w_{ij}I^{x^*}_{ij}}}, \label{eqn:aux_inc_xtilde}
\end{align}
where in~\eqref{eqn:aux_inc_xtilde} we use that $\tilde x_{ij} = x^*_{ij}$ for all $r_{ij}>\rho^*_j$ and zero otherwise. Define $\mu^*_{ij} = \ex{\frac{w_{ij}I^{x^*}_{ij}}{1+\sum_{i\in A_j}w_{ij}I^{x^*}_{ij}}}$ and $\mu^*_{0j} = 1-\sum_{i\in A_j}\mu^*_{ij}$. The following claim, whose proof is deferred to Appendix~\ref{app:proof_mustar_vs_pstar}, relates $\mu^*_{ij}$, $\mu^*_{0j}$, and $p^*_{ij}=\hat{w}_{ij}\cdot x^*_{ij}$.
\begin{claim}\label{clm:mustar_vs_pstar}
$\mu^*_{ij} \geq p^*_{ij} \cdot \mu^*_{0j}$, for every $i \in A_j$.
\end{claim}
From~\eqref{eqn:aux_inc_xtilde}, we observe that for every $j\in\S$
\begin{align}
  \ex{\finclusive_j(C^{\tilde x})} = \sum_{i\in A_j}r_{ij}\cdot \mu^*_{ij}  &= \rho^*_j\cdot\sum_{i\in A_j}\mu^*_{ij} + \sum_{i\in A_j}[r_{ij}-\rho^*_j]^+\cdot\mu^*_{ij} \notag\\
  &\geq \rho^*_j\cdot\sum_{i\in A_j}\mu^*_{ij} + \mu^*_{0j}\cdot \sum_{i\in A_j}p^*_{ij}\cdot [r_{ij}-\rho^*_j]^+\label{eqn:aux_inclusive_analysis1}\\
  & = \rho^*_j\cdot\sum_{i\in A_j}\mu^*_{ij} + \mu^*_{0j}\cdot \rho^*_j \label{eqn:aux_inclusive_analysis2}\\
  &= \rho^*_j \label{eqn:aux_inclusive_analysis3}
\end{align}
where in the second equality we use the fact that $A_j = \{i\in\C:\ r_{ij}>\rho^*_j\}$. Next, \eqref{eqn:aux_inclusive_analysis1} holds due to Claim~\ref{clm:mustar_vs_pstar}. Equality~\eqref{eqn:aux_inclusive_analysis2} follows since $\rho_j^*$ satisfies $\rho^*_j = \sum_{i \in \C} p^*_{ij} \cdot [r_{ij} - \rho^*_j]^+$ and $r_{ij}-\rho^*_j = 0$ for all $i\notin A_j$. Finally, we conclude~\eqref{eqn:aux_inclusive_analysis3} by noting that $\mu^*_{0j}+\sum_{i\in A_j}\mu^*_{ij} = 1$. Therefore, the first property follows by summing over $j\in\S$, 
\[
\revincP(\tilde x) = \sum_{j\in\S}\ex{\finclusive_j(C^{\tilde x})} \geq \sum_{j\in \S}\rho^*_j.
\]
We now prove the second property. Given an optimal solution $(x^*,y^*)$ of \eqref{eqn:customized_lp}, we have the following sequence of inequalities for every $j\in\S$
\begin{align}
\opt\eqref{eqn:customized_lp}=\sum_{i\in\C} r_{ij}\,y^*_{ij} &=\sum_{i\in\C} [\rho_j^*+r_{ij}-\rho_j^*]\cdot y^*_{ij} = \rho_j^*\cdot \sum_{i\in \C}y_{ij}^*+\sum_{i\in \C}[r_{ij}-\rho_j^*]\cdot y^*_{ij} \notag\\
&\leq \rho_{j}^* + \sum_{i\in \C}[r_{ij}-\rho_j^*]^+\cdot y^*_{ij} \label{eqn:aux_2t_bound1}\\
&\leq \rho_j^* + \sum_{i\in \C}[r_{ij}-\rho_{j}^*]^+\cdot \hat w_{ij}\cdot x_{ij}^*\, \label{eqn:aux_2t_bound2}\\
& =2\rho_j^*. \notag
\end{align}
In \eqref{eqn:aux_2t_bound1}, we use the fact that $y^*\in P^\S$, so $\sum_{i\in\C}y_{ij}^*\leq 1$, and $r_{ij}-\rho_j^*\leq \max\{r_{ij}-\rho_j^*,0\} = [r_{ij}-\rho_j^*]^+$. Next, in \eqref{eqn:aux_2t_bound2} we use that $y_{ij}^* = \hat{w}_{ij}\cdot x_{ij}^*$. The last equality uses that $\rho_j^*$ is the solution to the fixed point equation~\eqref{eqn:fixed_point}.
\end{proof}
\subsubsection{Related results.}\label{subsec:related_results}
As a consequence of Theorem~\ref{thm:main_algorithm}, we obtain the next corollary. 
\begin{corollary}\label{coro:main_customized}
For the customized model, there exists a randomized algorithm for computing a menu whose expected reward is $1/2$ of optimal.    
\end{corollary}
We note that this algorithm uses an optimal solution $x^*$ of \eqref{eqn:customized_lp} and its analysis is derived from Theorem~\ref{thm:main_algorithm}. A natural question is whether relaxation~\eqref{eqn:customized_lp} can lead to an approximation factor better than $1/2$. As the following lemma shows, this is not possible.
\begin{lemma}\label{lem:lowerbound_onehalf}
There exists an instance such that an optimal solution $(x^*,y^*)$ to \eqref{eqn:customized_lp}  satisfies $\frac{\revcustP(x^*)}{\opt\eqref{eqn:customized_lp}}\to\frac{1}{2}$ as the size of the market $n$ grows. In other words, no algorithm based on \eqref{eqn:customized_lp} can beat $1/2$.
\end{lemma}
We defer the proof to Appendix~\ref{app:lowerbound_onehalf}.
Finally, we establish the {reward gap} between the customized and the inclusive models, which is formally defined as
\[
\textsc{Reward-Gap} = \max_{\text{all instances }I}\frac{\optcust(I)}{\optinc(I)},
\]
where $\optcust(I)$ and $\optinc(I)$ are the optimal values for problem \eqref{eqn:problem_R} under the customized and inclusive model, respectively, for a given instance $I$. Thanks to Theorem~\ref{thm:main_algorithm} and an asymptotically worst-case instance, we show that the gap is 2.
\begin{theorem}\label{thm:revenue_gap}
The reward gap between the customized and inclusive models is 2. 
\end{theorem}
We defer the proof to Appendix~\ref{app:rewardgap}.

\section{Computational Study}\label{sec:computational_study}

In this section, we discuss our computational study under different types of instances. In Section~\ref{app:synthetic_study}, we assess the reward guarantees and running time performance of our proposed approximation algorithms as well as other benchmarks under synthetic instances. In Section~\ref{app:semisynthetic_study}, we consider a semi-synthetic large market instance to assess the reward performance under different levels of correlation between pairwise rewards and mutual attraction values.

\subsection{Synthetic Analysis}\label{app:synthetic_study}

In the following, we
utilize an extensive family of synthetic instances to
assess the performance of our proposed approximation algorithms for the customized and inclusive models in terms of their reward guarantees and running time efficiency. In particular,
we wish not only to demonstrate that our algorithmic methods perform in practice significantly better than their worst-case guarantees, but concurrently to show that the reward-wise performance  for the inclusive model does not fall much behind its analogous measure for the customized model. At the same time, we want to demonstrate that both methods are equally or substantially more efficient than the benchmarks.

\vspace{0.5em}

\noindent{\bf Experimental setup.} For this purpose, we focus  on two classes of synthetic instances:
\begin{itemize}
    \item {\it Heterogeneous-dense.} This class captures  the most general type of instances, where customers and suppliers
    are associated with arbitrary
    attraction values $u_{ij}$ and $w_{ij}$, respectively. Moreover, we consider a complete bipartite graph, where every customer-supplier pair $(i,j)$ is compatible.
    \item {\it Homogeneous-sparse.} For this class of instances, we adopt the more structured model of \citet{AshlagiKMSS22}, where values satisfy $u_{ij} = u_j$ and $w_{ij} = w_{j}$ for all $i\in \C, j\in \S$. Here, we consider a sparser bipartite graph where not every pair $(i,j)$ is compatible.
\end{itemize}
\vspace{0.5em}
For both families, we vary the number of customers  $n$ across 25, 50, 75, and 100, setting the number of suppliers to $m=n$. For each such pair, we generate 15 random instances where the attraction values $u_{ij}$ and $w_{ij}$ are generated according to the procedure detailed in Appendix~\ref{app:syntheticdetails}.

\vspace{0.5em}
\noindent{\bf Algorithms and Benchmarks.} 
For the customized model, we test our $(1/2)$-approximation (see Section~\ref{sec:customized}) as well as its corresponding upper bound~\eqref{eqn:customized_lp} referred to as $\algcust$ and $\lpcust$, respectively. For the inclusive model, we examine our $(1/2)$-approximation in this context (see Section~\ref{sec:customized})
 which we refer to as $\alginc$. 
 As previously explained, it is worth pointing out that~\eqref{eqn:customized_lp} provides an upper bound for the inclusive model as well. As comparative benchmarks, we tested the next three alternatives:
\begin{itemize}
\item To examine whether trivial solutions perform poorly, we test the full menu in which all suppliers appear in the menu offered. This naive approach is denoted by ``All''.
\item We also implemented the $(1/2-\epsilon)$-approximation and LP-based upper bound by~\citet{ahmadnejadsaein2025adaptive} for the customized model, referred to as $\algatar$ and $\lpatar$, respectively. While their approach was proposed for a different model where the platform can adaptively design the assortment for each customer, their solution concept is static (i.e., all menus are decided in advance), meaning that $\algatar$ applies to our setting. In addition, their approach is centered around a linear program $\lpatar$ with exponentially many variables, which upper bounds our customized model. To speed up the solution time of $\lpatar$, we solve its corresponding separation problem using an MIQCP formulation with an optimality-gap tolerance  of $0.05$ along with a cutoff value for pruning and a best objective bound as an early stopping condition. Moreover, we limit the number of new columns added to the master problem, meaning that the $\lpatar$ results below should be interpreted as a close approximation to the true optimal value. 
\item We implemented a greedy approach for the customized and inclusive models, inspired by the arbitrary-order greedy with simulated choices, studied by~\citet{TorricoCL22}, referred to  as $\greedycust$ and $\greedyinc$. Specifically, this algorithm receives as input an arbitrary order over the customers $i_1,\ldots,i_n$. For each customer $i_t$, it computes an assortment $M_{i_t}$ by solving the single-customer assortment problem 
\[\max_{S\subseteq\S}\sum_{j\in S} \left[f_j(C^{t-1}_j\cup\{i_t\})-f_j(C^{t-1}_j)\right] \cdot\pic_{i_t}(j,S)
\]
where $f_j(C^{t-1}_j\cup\{i_t\})-f_j(C^{t-1}_j)$
represents the marginal reward obtained by adding customer $i_t$ to the ``current'' backlog $C_j^{t-1}$ of supplier $j$. Once $M_{i_t}$ is obtained, this method simulates the choice of customer $i_t$ and updates the corresponding backlog. Namely, if $i_t$ chooses $j$, then $C_j^t = C^{t-1}_j\cup\{i_t\}$. For $\greedycust$ and $\greedyinc$, we utilize $f_j = \fcustomized_j$ and $f_j = \finclusive_j$, respectively.
\end{itemize}

\vspace{0.5em}
\noindent{\bf Simulations.} We remind readers that the objective function in either model is the expected total reward computed over the customers' random choices. Therefore, for a given menu, we estimate its expected reward by averaging 100 samples. Formally, for a fractional solution $x\in P^\C$, one sample corresponds to drawing the random variables $I^x_{ij} \sim \mybern(x_{ij})$ for all $(i,j)\in\E$. For a discrete solution, i.e., a menu $M = (M_i)_{i \in \C}$, one sample corresponds to every customer $i$ selecting a supplier $j$ or the outside option according to her choice model $\pic_i$.

\vspace{0.5em}

\noindent{\bf Results and Interpretation.} 
For homogeneous instances, the reward guarantees and running time performance of our algorithms, as well as those of the aforementioned benchmarks, are presented in Tables~\ref{tab:obj_homogeneous} and~\ref{tab:time_homogeneous}. In particular, the first part of Table~\ref{tab:obj_homogeneous} shows the optimal objective values of the upper bounds for the customized model. In the next panel, Table~\ref{tab:obj_homogeneous} presents the reward guarantees of each method as the ratio between its objective value against each upper bound, $\mathrm{ALG}/\lpcust$ and $\mathrm{ALG}/\lpatar$. In the last panel, Table~\ref{tab:obj_homogeneous} shows the ratios for the inclusive model. 
 Table~\ref{tab:obj_homogeneous} demonstrates,  our upper bound $\lpcust$ nearly matches $\lpatar$. This finding makes our polynomial-size relaxation more appealing for practical purposes as it can be solved efficiently without needing column generation.  
 Regarding our approach for the customized model, the ratio between $\algcust$ and $\lpcust$ decreases as the number of customers grows, with  0.845 for $n=25$ and 0.768 for $n=100$; however, this empirical performance is still much better than our worst-case $(1/2)$-approximation. 
With respect to the additional benchmarks,
$\algatar$ performs similarly to $\algcust$ and $\greedycust$ performs 2--3\% worse. Lastly, the trivial solution All performs far worse than all other methods.
In terms of efficiency, Table~\ref{tab:time_homogeneous} shows that $\algcust$ is significantly more appealing as its average running time is less than 0.13 seconds even for $n=100$, as opposed to $\algatar$ and $\greedycust$, which require about 5 and 1 minutes for such instances, respectively.  For the inclusive model, we note that the reward performance of $\algatar$ is the best among all inclusive methods, although $\alginc$ and $\greedyinc$ are both very competitive options. Still, from a practical point of view, $\alginc$ and $\greedyinc$ have the best value-efficiency tradeoff as their average running times are much better than that of $\algatar$. That said, we do not know whether $\greedyinc$ guarantees any constant-factor approximation. 

\begin{table}[htbp]
\caption{Reward guarantees under the customized and inclusive models on homogeneous instances.}
\label{tab:obj_homogeneous}
\centerline{
\scalebox{0.75}{
\begin{tabular}{c cc cccc |  cccc}
\toprule
& \multicolumn{6}{c}{Customized} & \multicolumn{4}{c}{Inclusive} \\
\cmidrule(lr){2-7}\cmidrule(lr){8-11}
& \multicolumn{2}{c}{UB} & \multicolumn{4}{c}{Ratios \quad $\frac{\mathrm{ALG}}{\lpcust},\;\frac{\mathrm{ALG}}{\lpatar}$}  & \multicolumn{4}{c}{Ratios \quad $\frac{\mathrm{ALG}}{\lpcust}$} \\
\cmidrule(lr){2-3}\cmidrule(lr){4-7}\cmidrule(lr){8-11}
$n=m$ & $\lpcust$ & $\lpatar$ & $\algcust$ & $\algatar$ & $\greedycust$ & All  & $\alginc$ & $\algatar$ & $\greedyinc$ & All \\
\midrule
25 & 24.92 & 24.45 & 0.845,\;0.862 & 0.843,\;0.861 & 0.815,\;0.831 & 0.502,\;0.512  & 0.841 & 0.841 & 0.800 & 0.491 \\
50 & 60.83 & 59.73 & 0.795,\;0.810 & 0.816,\;0.831 & 0.774,\;0.789 & 0.438,\;0.447  & 0.797 & 0.818 & 0.760 & 0.424 \\
75 & 107.11 & 104.52 & 0.759,\;0.778 & 0.787,\;0.808 & 0.725,\;0.743 & 0.389,\;0.399  & 0.767 & 0.789 & 0.718 & 0.373 \\
100 & 143.20 & 140.66 & 0.768,\;0.782 & 0.790,\;0.805 & 0.736,\;0.749 & 0.393,\;0.400  & 0.773 & 0.793 & 0.739 & 0.380 \\
\bottomrule
\end{tabular}}}
\end{table}
\begin{table}[htbp!]
\caption{Running times (seconds) on homogeneous instances.}
\label{tab:time_homogeneous}
\centerline{
\scalebox{0.7}{\begin{tabular}{c ccc | cc}
\toprule
& \multicolumn{3}{c}{Customized} & \multicolumn{2}{c}{Inclusive} \\
\cmidrule(lr){2-4}\cmidrule(lr){5-6}
$n=m$ & $\algcust$ & $\algatar$ & $\greedycust$ & $\alginc$ & $\greedyinc$ \\
\midrule
25 & 0.00 & 21.31 & 4.43 & 0.64 & 0.24 \\
50 & 0.02 & 84.64 & 18.09 & 3.13 & 0.77 \\
75 & 0.08 & 176.33 & 40.47 & 8.74 & 1.58 \\
100 & 0.13 & 306.79 & 69.12 & 20.40 & 2.53 \\
\bottomrule
\end{tabular}}}
\end{table}

For heterogeneous instances, the reward guarantees and running times of our  algorithms as well as those of the additional benchmarks can be found in Tables~\ref{tab:obj_heterogeneous} and~\ref{tab:time_heterogeneous}. First, with the exception of the trivial solution All, all methods tested improve their performance relative to the homogeneous instances, in terms of both reward and efficiency. This trend is expected as the platform can effectively tailor the menu of each customer in a more diverse market, which is not the case with All. Second, we note that $\algatar$ achieves slightly higher reward than $\algcust$ in the customized model and than $\alginc$ in the inclusive setting. However, this minor improvement comes at the expense of impractical running time, on the order of several minutes as demonstrated in Table~\ref{tab:time_heterogeneous}. The remaining  observations are analogous to our conclusions for homogeneous instances.
\begin{table}[htbp!]
\caption{Reward guarantees under the customized and inclusive models on heterogeneous instances.}
\label{tab:obj_heterogeneous}
\centerline{
\scalebox{0.75}{\begin{tabular}{c cc cccc |  cccc}
\toprule
& \multicolumn{6}{c}{Customized} & \multicolumn{4}{c}{Inclusive} \\
\cmidrule(lr){2-7}\cmidrule(lr){8-11}
& \multicolumn{2}{c}{UB} & \multicolumn{4}{c}{Ratios \quad $\frac{\mathrm{ALG}}{\lpcust},\;\frac{\mathrm{ALG}}{\lpatar}$}  & \multicolumn{4}{c}{Ratios \quad $\frac{\mathrm{ALG}}{\lpcust}$} \\
\cmidrule(lr){2-3}\cmidrule(lr){4-7}\cmidrule(lr){8-11}
$n=m$ & $\lpcust$ & $\lpatar$ & $\algcust$ & $\algatar$ & $\greedycust$ & All  & $\alginc$ & $\algatar$ & $\greedyinc$ & All \\
\midrule
25 & 10.15 & 9.92 & 0.843,\;0.862 & 0.853,\;0.874 & 0.831,\;0.852 & 0.415,\;0.424 & 0.834 & 0.846 & 0.833 & 0.404 \\
50 & 24.54 & 24.10 & 0.823,\;0.839 & 0.828,\;0.844 & 0.796,\;0.811 & 0.363,\;0.371  & 0.814 & 0.821 & 0.781 & 0.353 \\
75 & 39.77 & 39.18 & 0.811,\;0.823 & 0.823,\;0.835 & 0.784,\;0.796 & 0.334,\;0.340 & 0.795 & 0.813 & 0.774 & 0.327 \\
100 & 56.11 & 55.40 & 0.814,\;0.825 & 0.824,\;0.836 & 0.782,\;0.792 & 0.319,\;0.324  & 0.799 & 0.815 & 0.771 & 0.310 \\
\bottomrule
\end{tabular}}}
\end{table}

 One natural question is whether personalization is practically effective, in the sense that $\algcust$ would perform much better than $\alginc$. Quite surprisingly this is the case in neither  homogeneous nor heterogeneous instances. For example, in homogeneous instances with $n=25$ customers, $\algcust$ achieves an expected total reward of $0.845\cdot 24.92\approx 21.05$ while $\alginc$ garners  on average $0.841\cdot 24.92 \approx20.95$. Analogous results can be concluded for other instances, indicating that with respect to the scenarios tested, personalization can only slightly improve the expected reward. We remark that in certain cases, $\alginc$ appears to perform better than $\algcust$, mainly due to the variance when estimating their objective value.

\begin{table}[htbp!]
\caption{Running times (seconds) on heterogeneous instances.}
\label{tab:time_heterogeneous}
\centerline{
\scalebox{0.75}{\begin{tabular}{c ccc | cc}
\toprule
& \multicolumn{3}{c}{Customized} & \multicolumn{2}{c}{Inclusive} \\
\cmidrule(lr){2-4}\cmidrule(lr){5-6}
$n=m$ & $\algcust$ & $\algatar$ & $\greedycust$ & $\alginc$ & $\greedyinc$ \\
\midrule
25 & 0.00 & 16.70 & 4.76 & 0.60 & 0.25 \\
50 & 0.02 & 43.83 & 18.08 & 3.10 & 0.75 \\
75 & 0.03 & 107.51 & 39.86 & 8.85 & 1.55 \\
100 & 0.06 & 243.43 & 69.45 & 20.36 & 2.54 \\
\bottomrule
\end{tabular}}}
\end{table}

\subsection{Reward-Preference Correlation Study}\label{app:semisynthetic_study}

In this study, we evaluate our algorithmic methods along with additional benchmarks on a semi-synthetic large market instance, calibrated to real labor-market data. Our main goal is to assess the reward performance of these methods under different levels of correlation between pairwise rewards and mutual preference values. Intuitively, when rewards align with attraction weights, reward-maximizing menus coincide with match-maximizing ones (thereby defining an easy regime), whereas negative correlation forces the platform to trade match probability against reward (capturing the {worst-case} regime). 

\textbf{Experimental setup.} We construct a two-sided labor market with
$n = 500$ customers (job seekers) and $m = 100$ suppliers
((employer,seniority-level) openings) using three public datasets: a LinkedIn job
postings corpus, with views and application counts per posting~\citep{kon2024linkedin}, a data-science
salary survey~\citep{chaki2025dssalaries}, and a Glassdoor employer ratings dataset~\citep{gauthier2021glassdoor}, restricted to data-science
roles. In what follows, we explain how the three parameter groups $(u,w,r)$ are obtained, deferring additional details
to Appendix~\ref{app:semidetails}:
\begin{itemize}
\item \emph{Customer weights $u_{ij}$ are estimated.} Since only aggregate
counts are observed, job-seekers are grouped into $T = 12$ types (seniority level
$\times$ skill cluster), and each type's preference weights are estimated from the
per-posting apply rates by an aggregate-share logit inversion in the spirit of~\citet{berry1994estimating}. Under the MNL choice model, for a job-seeker type $t$, the probability of applying when offered menu $M_t = \{j\}$ equal $u_{tj}/(1+u_{tj})$, which leads us to regress  empirical log-odds on posting covariates (salary, employer rating, remote match, skill match) via  weighted least squares method. The final $n=500$ job-seekers are then sampled from the estimated type mixture with random coefficients, analogous to a mixed MNL choice model.
\item \emph{Supplier weights $w_{ij}$ are semi-synthetic.} Recruiter choices
are not observable in public data, and we therefore set supplier weights such that employers softly prefer higher-quality candidates that match the skill and seniority level requirements. Covariates are observable and their coefficients are some assumed values.
\item \emph{Rewards $r_{ij}$ are semi-synthetic and correlated with attraction values.}
In this study, rewards represent the fee that the platform receives per match. Each reward value  is drawn
from a log-normal marginal scaled to a $20\%$ commission on the median
posted salary and coupled to the attraction signal $\log(u_{ij} w_{ij})$
through Gaussian noise with a correlation factor
$\rho \in \{-0.9, -0.6, \allowbreak -0.3, 0, 0.3, 0.6, 0.9\}$. By
construction, the {marginal} reward distribution is identical for every $\rho$, $u$, and $w$, and the sampled individuals are fixed across the range of $\rho$-values, so any
performance difference is an outcome only of the correlation.
\end{itemize}

\textbf{Algorithms and benchmarks.} As in Section~\ref{app:synthetic_study}, we tested our $(1/2)$-approximation for the customized model and its upper bound ($\algcust$ and $\lpcust$), our $(1/2)$-approximation for the inclusive model ($\alginc$), and the simulated-greedy benchmarks $\greedycust$ and $\greedyinc$. We did not test $\algatar$ due to its prohibitive running time for an instance with $n=500$ customers. To isolate the value of accounting for pairwise rewards,
we also consider a \emph{reward-oblivious} algorithm in the customized model
($\algcustobv$) that corresponds to solving the same linear program with all reward values set to $1$, whose resulting fractional solution is then evaluated under the true
reward objective. In this case, the reward-oblivious inclusive method coincides with its customized counterpart. As before, expected rewards are estimated over $100$ samples of the sequential matching process, with the reward guarantees of each method reported relative to $\lpcust$.

\begin{table}[htbp]
\centering
\caption{Reward guarantees under the semi-synthetic instance
per correlation factor $\rho$. }
\label{tab:semi}
\scalebox{0.75}{\begin{tabular}{ccccc|cc}
\toprule
& \multicolumn{4}{c}{Customized} & \multicolumn{2}{c}{Inclusive} \\
\cmidrule(lr){2-5}\cmidrule(lr){6-7}
& UB & \multicolumn{3}{c}{Ratios against $\lpcust$}
& \multicolumn{2}{c}{Ratios against $\lpcust$} \\
\cmidrule(lr){2-2}\cmidrule(lr){3-5}\cmidrule(lr){6-7}
$\rho$ & $\lpcust$ & $\algcust$ & $\algcustobv$ & $\greedycust$
         & $\alginc$   & $\greedyinc$ \\
\midrule
$-0.9$ & $5614.4$ & $0.575$ & $0.282$ & $0.567$  & $0.575$  & $0.536$ \\
$-0.6$ & $7107.2$ & $0.607$ & $0.251$ & $0.593$  & $0.608$  & $0.556$ \\
$-0.3$ & $8239.4$ & $0.623$ & $0.231$ & $0.619$  & $0.622$  & $0.578$ \\
$\phantom{-}0.0$ & $9237.5$ & $0.651$ & $0.223$ & $0.666$  & $0.642$  & $0.637$ \\
$\phantom{-}0.3$ & $9489.9$ & $0.660$ & $0.236$ & $0.643$ & $0.656$ & $0.576$ \\
$\phantom{-}0.6$ & $9010.2$ & $0.658$ & $0.262$ & $0.673$  & $0.658$  & $0.595$ \\
$\phantom{-}0.9$ & $7311.3$ & $0.656$ & $0.333$ & $0.686$  & $0.651$  & $0.659$ \\
\bottomrule
\end{tabular}}
\end{table}

\textbf{Results.} Table~\ref{tab:semi} reports the optimal reward of $\lpcust$ and the reward guarantees computed as the ratio between the corresponding method and $\lpcust$, where two findings particularly stand out. First, every ratio generally increases with $\rho$: $\algcust$ increases from $0.575$ at $\rho = -0.9$ to $0.656$ at $\rho = 0.9$, and $\greedycust$ from $0.567$ to $0.686$. This phenomenon confirms the intuition described above: Aligning rewards with preference weights gradually shifts instances from the worst-case toward the easy regime. Despite this change in reward performance, even at $\rho = -0.9$, our methods perform significantly better than their worst-case guarantees. Second, the achievable reward itself is {single-peaked} rather than monotone: $\lpcust$ is maximized at a correlation level of $\rho \approx 0.3$. To interpret this result, we note that the marginal reward distribution is the same across different $\rho$-values, which means that the group of high-reward pairs has the same size for different $\rho$. However, extreme positive correlation concentrates high rewards on pairs that were already highly attractive and would be matched anyway, which is somewhat wasteful. Instead, under moderate alignment, the same pool of high reward values is spread across pairs, leading to a potentially higher expected total reward. The ratios nevertheless keep improving, since the value of $\lpcust$ shrinks along with the obtained reward. Finally, mirroring the synthetic instances in Appendix~\ref{app:synthetic_study}, personalization is only mildly beneficial: $\alginc$ attains a comparable expected reward to $\algcust$ across the range of $\rho$-values, and the greedy variants are strong empirical performers in both models.

\textbf{The value of reward-awareness.} At least conceptually, the $\algcustobv$ column of Table~\ref{tab:semi} isolates the extent to which the pairwise rewards drive the design of each menu relative to preference weights. Since this oblivious policy ignores pairwise rewards, its computed set of menus is
identical at every correlation level $\rho$, and its ratio against $\lpcust$ varies very little. However,
its evaluated reward increases roughly monotonically with $\rho$ 
(e.g., collecting $0.282\cdot 5614.4\approx 1583$ at $\rho=-0.9$
and $0.333\cdot 7311.3\approx 2435$ at $\rho=0.9$), since high-reward
pairs tend to coincide with high-attraction pairs, the latter being the main input in $\algcustobv$.
Overall, a reward-oblivious policy captures only $22$--$33\%$ of the LP upper
bound.

Along these lines, a cleaner comparison is between the oblivious and reward-aware
approaches, summarized by the relative gap $(\algcust-\algcustobv)/\algcust$
in Table~\ref{tab:value}. Reward-awareness roughly doubles the collected
reward throughout, but its value is \emph{not} monotone in
$\rho$, with a peak in the middle. The observed gap is largest when rewards are essentially
uncorrelated with preference weights ($\rho\approx 0$), where the reward-oblivious
algorithm recovers only $\frac{2060.0}{6013.6}\approx 34\%$ of the reward-aware reward (a relative gap of $0.657$),
and it is smaller at {both} extremes, where the oblivious method recovers about
$49$--$51\%$. At $\rho=+0.9$ the gap narrows since maximizing matches nearly
coincides with maximizing reward. At $\rho=-0.9$ the gap narrows as well for the
opposite reason: the high rewards now sit on low-attraction pairs that rarely
match, so {even the reward-aware algorithm cannot collect them}. In other  words, being reward-aware buys little when the valuable pairs are essentially unmatchable. Reward-awareness is thus most valuable precisely when reward and preference are decoupled, while remaining very substantial across the entire range.

\begin{table}[t]
\centering
\caption{Reward values and relative gaps per correlation factor $\rho$.}
\label{tab:value}
\scalebox{0.85}{\begin{tabular}{cccc}
\toprule
$\rho$ & $\algcust$ & $\algcustobv$ & $\frac{\algcust-\algcustobv}{\algcust}$ \\
\midrule
$-0.9$ & $3228.3$ & $1583.3$ & $0.510$ \\
$-0.6$ & $4314.1$ & $1783.9$ & $0.586$ \\
$-0.3$ & $5133.1$ & $1903.3$ & $0.629$ \\
$\phantom{-}0.0$ & $6013.6$ & $2060.0$ & $0.657$ \\
$\phantom{-}0.3$ & $6263.3$ & $2239.6$ & $0.642$ \\
$\phantom{-}0.6$ & $5928.7$ & $2360.7$ & $0.602$ \\
$\phantom{-}0.9$ & $4796.2$ & $2434.7$ & $0.492$ \\
\bottomrule
\end{tabular}}
\end{table}
\section{Conclusions}\label{sec:conclusions}
We conclude this work by highlighting a number of particularly interesting open questions for future research. Specifically, we bring up several technical insights that may be useful in improving our approximation guarantees. 

\paragraph{Improved or tight guarantees?} As mentioned in Section~\ref{sec:intro}, the lack of submodularity and monotonicity of our reward functions gives rise to unique algorithmic obstacles and analytic challenges. While we have shown that these issues can be circumvented via quite a few novel developments, it would be interesting to investigate whether our approximation guarantees are tight. Lemma~\ref{lem:lowerbound_onehalf} shows that \eqref{eqn:customized_lp} cannot lead to an approximation factor better than $1/2$. 
The question is whether this constant can be improved via a different approach while avoiding an extremely involved  analysis, or whether no algorithm can do better. 

\paragraph{Stronger combinatorial guarantees.} We briefly mention that the purely combinatorial guarantees of Section~\ref{sec:purely-combinatorial}, obtained by an elementary greedy approximation of our proxy model, can be improved without resorting to mathematical programming. Indeed, as Lemma~\ref{lem:comb-submodular} shows, our equivalent formulation involves a monotone submodular function $F : 2^{\E} \to \bbR_+$, to be optimized subject to a partition matroid constraint. The non-oblivious local-search framework of \citet{Filmus14} attains a $(1-\frac{1}{e})$-approximation in this setting. More recently, \citet{BuchbinderF24} obtained a deterministic $(1-\frac{1}{e}-\eps)$-approximation for monotone submodular maximization subject to an arbitrary matroid constraint, while \citet{RozenmanKSS26} developed an alternative approach based on a Poisson process, attaining a similar guarantee without relying on discretization or rounding. Although our corresponding ground set is exponentially large, its special MNL-based structure allows the required local-search marginal optimization to be implicitly carried out through polynomial-time assortment optimization oracles. Combining these ideas with the proxy-to-inclusive conversion of Section~\ref{subsec:comb-conversion} yields a purely combinatorial approach for both models with approximation ratio $\frac{1 }{ 2} \cdot (1 - \frac{ 1 }{ e })$. Since establishing this implementation requires a rather lengthy non-oblivious local-search analysis while being orthogonal to our main ideas, we do not elaborate on its finer details. However, the question remains open: Does there exist a simple combinatorial approach with a guarantee better than $1/4$?



\addcontentsline{toc}{section}{Bibliography}
\bibliographystyle{plainnat}
\bibliography{BIB-Seq-Matching}

\begin{thebibliography}{52}
\providecommand{\natexlab}[1]{#1}
\providecommand{\url}[1]{\texttt{#1}}
\expandafter\ifx\csname urlstyle\endcsname\relax
  \providecommand{\doi}[1]{doi: #1}\else
  \providecommand{\doi}{doi: \begingroup \urlstyle{rm}\Url}\fi

\bibitem[Adamczyk(2011)]{Adamczyk11}
Marek Adamczyk.
\newblock Improved analysis of the greedy algorithm for stochastic matching.
\newblock \emph{Information Processing Letters}, 111\penalty0 (15):\penalty0
  731--737, 2011.

\bibitem[Adamczyk et~al.(2015)Adamczyk, Grandoni, and Mukherjee]{AdamczykGM15}
Marek Adamczyk, Fabrizio Grandoni, and Joydeep Mukherjee.
\newblock Improved approximation algorithms for stochastic matching.
\newblock In \emph{Proceedings of the 23rd Annual European Symposium on
  Algorithms}, pages 1--12, 2015.

\bibitem[Ahmadnejadsaein and {El Housni}(2025)]{ahmadnejadsaein2025adaptive}
Mohammadreza Ahmadnejadsaein and Omar {El Housni}.
\newblock Adaptive two-sided assortment optimization: Revenue maximization.
\newblock \emph{arXiv preprint arXiv:2507.04156}, 2025.

\bibitem[Ahmed et~al.(2022)Ahmed, Sohoni, and Bandi]{AhmedSB22}
Asrar Ahmed, Milind~G. Sohoni, and Chaithanya Bandi.
\newblock Parameterized approximations for the two-sided assortment
  optimization.
\newblock \emph{Operation Research Letters}, 50\penalty0 (4):\penalty0
  399--406, 2022.

\bibitem[Aouad and Saban(2023)]{AouadS23}
Ali Aouad and Daniela Saban.
\newblock Online {{assortment optimization}} for {{two-sided matching
  platforms}}.
\newblock \emph{Management Science}, 69\penalty0 (4):\penalty0 2069--2087,
  2023.

\bibitem[Aouad et~al.(2018)Aouad, Farias, Levi, and Segev]{AouadFRS18}
Ali Aouad, Vivek Farias, Retsef Levi, and Danny Segev.
\newblock The approximability of assortment optimization under ranking
  preferences.
\newblock \emph{Operations Research}, 66\penalty0 (6):\penalty0 1661--1669,
  2018.

\bibitem[Ashlagi et~al.(2022)Ashlagi, Krishnaswamy, Makhijani, Saban, and
  Shiragur]{AshlagiKMSS22}
Itai Ashlagi, Anilesh~K. Krishnaswamy, Rahul Makhijani, Daniela Saban, and
  Kirankumar Shiragur.
\newblock Assortment planning for two-sided sequential matching markets.
\newblock \emph{Operations Research}, 70\penalty0 (5):\penalty0 2784--2803,
  2022.

\bibitem[Bansal et~al.(2012)Bansal, Gupta, Li, Mestre, Nagarajan, and
  Rudra]{BansalGLMNR12}
Nikhil Bansal, Anupam Gupta, Jian Li, Juli{\'a}n Mestre, Viswanath Nagarajan,
  and Atri Rudra.
\newblock When {LP} is the cure for your matching woes: Improved bounds for
  stochastic matchings.
\newblock \emph{Algorithmica}, 63:\penalty0 733--762, 2012.

\bibitem[Baveja et~al.(2018)Baveja, Chavan, Nikiforov, Srinivasan, and
  Xu]{BavejaCNSX18}
Alok Baveja, Amit Chavan, Andrei Nikiforov, Aravind Srinivasan, and Pan Xu.
\newblock Improved bounds in stochastic matching and optimization.
\newblock \emph{Algorithmica}, 80:\penalty0 3225--3252, 2018.

\bibitem[Berry(1994)]{berry1994estimating}
Steven~T. Berry.
\newblock Estimating discrete-choice models of product differentiation.
\newblock \emph{The RAND Journal of Economics}, 25\penalty0 (2):\penalty0
  242--262, 1994.

\bibitem[Blanchet et~al.(2016)Blanchet, Gallego, and Goyal]{BlanchetGG16}
Jose Blanchet, Guillermo Gallego, and Vineet Goyal.
\newblock A {M}arkov chain approximation to choice modeling.
\newblock \emph{Operations Research}, 64\penalty0 (4):\penalty0 886--905, 2016.

\bibitem[Borodin and MacRury(2025)]{BorodinM25}
Allan Borodin and Calum MacRury.
\newblock Online bipartite matching in the probe-commit model.
\newblock \emph{Mathematical Programming (forthcoming)}, pages 1--54, 2025.

\bibitem[Brubach et~al.(2024)Brubach, Grammel, Ma, and
  Srinivasan]{BrubachGMS21}
Brian Brubach, Nathaniel Grammel, Will Ma, and Aravind Srinivasan.
\newblock Improved guarantees for offline stochastic matching via new ordered
  contention resolution schemes.
\newblock \emph{Mathematics of Operations Research (forthcoming)}, 2024.

\bibitem[Buchbinder and Feldman(2018)]{BuchbinderF18Survey}
Niv Buchbinder and Moran Feldman.
\newblock Submodular functions maximization problems.
\newblock In Teofilo~F. Gonzalez, editor, \emph{Handbook of Approximation
  Algorithms and Metaheuristics}, pages 753--788. Chapman and Hall/CRC, 2018.

\bibitem[Buchbinder and Feldman(2024)]{BuchbinderF24}
Niv Buchbinder and Moran Feldman.
\newblock Deterministic algorithm and faster algorithm for submodular
  maximization subject to a matroid constraint.
\newblock In \emph{Proceedings of the 65th {IEEE} Annual Symposium on
  Foundations of Computer Science}, pages 700--712, 2024.

\bibitem[C{\u a}linescu et~al.(2011)C{\u a}linescu, Chekuri, P{\'a}l, and
  Vondr{\'a}k]{CalinescuCPV11}
Gruia C{\u a}linescu, Chandra Chekuri, Martin P{\'a}l, and Jan Vondr{\'a}k.
\newblock Maximizing a monotone submodular function subject to a matroid
  constraint.
\newblock \emph{SIAM Journal on Computing}, 40\penalty0 (6):\penalty0
  1740--1766, 2011.

\bibitem[Chaki(2025)]{chaki2025dssalaries}
Arnab Chaki.
\newblock Data science salaries 2025.
\newblock Kaggle dataset, 2025.
\newblock
  \url{https://www.kaggle.com/datasets/arnabchaki/data-science-salaries-2025},
  accessed August 2026.

\bibitem[Chen et~al.(2009)Chen, Immorlica, Karlin, Mahdian, and
  Rudra]{ChenIKMR09}
Ning Chen, Nicole Immorlica, Anna~R. Karlin, Mohammad Mahdian, and Atri Rudra.
\newblock Approximating matches made in heaven.
\newblock In \emph{Proceedings of the 36th International Colloquium on
  Automata, Languages and Programming}, pages 266--278, 2009.

\bibitem[Costello et~al.(2012)Costello, Tetali, and Tripathi]{CostelloTT12}
Kevin~P. Costello, Prasad Tetali, and Pushkar Tripathi.
\newblock Stochastic matching with commitment.
\newblock In \emph{Proceedings of the 39th International Colloquium on
  Automata, Languages, and Programming}, pages 822--833, 2012.

\bibitem[Davis et~al.(2014)Davis, Gallego, and Topaloglu]{DavisGT14}
James~M. Davis, Guillermo Gallego, and Huseyin Topaloglu.
\newblock Assortment optimization under variants of the nested logit model.
\newblock \emph{Operations Research}, 62\penalty0 (2):\penalty0 250--273, 2014.

\bibitem[Derakhshan and Farhadi(2023)]{DerakhshanF23}
Mahsa Derakhshan and Alireza Farhadi.
\newblock Beating {($1-1/e$)}-approximation for weighted stochastic matching.
\newblock In \emph{Proceedings of the 2023 Annual ACM-SIAM Symposium on
  Discrete Algorithms}, pages 1931--1961, 2023.

\bibitem[D{\'e}sir et~al.(2020)D{\'e}sir, Goyal, Segev, and Ye]{DesirGSY20}
Antoine D{\'e}sir, Vineet Goyal, Danny Segev, and Chun Ye.
\newblock Constrained assortment optimization under the {M}arkov chain-based
  choice model.
\newblock \emph{Management Science}, 66\penalty0 (2):\penalty0 698--721, 2020.

\bibitem[D{\'e}sir et~al.(2022)D{\'e}sir, Goyal, and Zhang]{DesirGZ22}
Antoine D{\'e}sir, Vineet Goyal, and Jiawei Zhang.
\newblock Capacitated assortment optimization: Hardness and approximation.
\newblock \emph{Operations Research}, 70\penalty0 (2):\penalty0 893--904, 2022.

\bibitem[El~Housni and Topaloglu(2023)]{el2023joint}
Omar El~Housni and Huseyin Topaloglu.
\newblock Joint assortment optimization and customization under a mixture of
  multinomial logit models: On the value of personalized assortments.
\newblock \emph{Operations research}, 71\penalty0 (4):\penalty0 1197--1215,
  2023.

\bibitem[{El Housni} et~al.(2024){El Housni}, Hennebelle, and
  Torrico]{ElHousniHT24}
Omar {El Housni}, Ulysse Hennebelle, and Alfredo Torrico.
\newblock Two-sided assortment optimization: Adaptivity gaps and approximation
  algorithms, 2024.
\newblock Working paper. Available online as arXiv report 2403.08929.
  Preliminary version in {\em Proceedings of the 25th International Conference
  on Integer Programming and Combinatorial Optimization}, 2024.

\bibitem[Farias et~al.(2013)Farias, Jagabathula, and Shah]{FariasJS13}
Vivek~F. Farias, Srikanth Jagabathula, and Devavrat Shah.
\newblock A nonparametric approach to modeling choice with limited data.
\newblock \emph{Management Science}, 59\penalty0 (2):\penalty0 305--322, 2013.

\bibitem[Feldman and Topaloglu(2017)]{FeldmanT17}
Jacob~B. Feldman and Huseyin Topaloglu.
\newblock Revenue management under the markov chain choice model.
\newblock \emph{Operations Research}, 65\penalty0 (5):\penalty0 1322--1342,
  2017.

\bibitem[Feng et~al.(2022)Feng, Niazadeh, and Saberi]{FengNS22}
Yiding Feng, Rad Niazadeh, and Amin Saberi.
\newblock Near-optimal bayesian online assortment of reusable resources.
\newblock In \emph{Proceedings of the 23rd ACM Conference on Economics and
  Computation}, pages 964--965, 2022.

\bibitem[Filmus and Ward(2014)]{Filmus14}
Yuval Filmus and Justin Ward.
\newblock Monotone submodular maximization over a matroid via non-oblivious
  local search.
\newblock \emph{SIAM Journal on Computing}, 43\penalty0 (2):\penalty0 514--542,
  2014.

\bibitem[Fu et~al.(2021)Fu, Tang, Wu, Wu, and Zhang]{FuTWWZ21}
Hu~Fu, Zhihao~Gavin Tang, Hongxun Wu, Jinzhao Wu, and Qianfan Zhang.
\newblock Random order vertex arrival contention resolution schemes for
  matching, with applications.
\newblock In \emph{Proceedings of the 48th International Colloquium on
  Automata, Languages, and Programming}, pages 68:1--68:20, 2021.

\bibitem[Gallego and Topaloglu(2014)]{GallegoT14}
Guillermo Gallego and Huseyin Topaloglu.
\newblock Constrained assortment optimization for the nested logit model.
\newblock \emph{Management Science}, 60\penalty0 (10):\penalty0 2583--2601,
  2014.

\bibitem[Gallego and Topaloglu(2019)]{GallegoT19}
Guillermo Gallego and Huseyin Topaloglu.
\newblock \emph{Revenue Management and Pricing Analytics}.
\newblock {Springer}, 2019.

\bibitem[Gallego et~al.(2015)Gallego, Ratliff, and Shebalov]{GallegoRS15}
Guillermo Gallego, Richard Ratliff, and Sergey Shebalov.
\newblock A general attraction model and sales-based linear program for network
  revenue management under customer choice.
\newblock \emph{Operations Research}, 63\penalty0 (1):\penalty0 212--232, 2015.

\bibitem[Gamlath et~al.(2019)Gamlath, Kale, and Svensson]{GamlathKS19}
Buddhima Gamlath, Sagar Kale, and Ola Svensson.
\newblock Beating greedy for stochastic bipartite matching.
\newblock In \emph{Proceedings of the 30th Annual ACM-SIAM Symposium on
  Discrete Algorithms}, pages 2841--2854, 2019.

\bibitem[{Ganz Rozenman} et~al.(2026){Ganz Rozenman}, Kulik, Schwartz, and
  Singh]{RozenmanKSS26}
Amit {Ganz Rozenman}, Ariel Kulik, Roy Schwartz, and Mohit Singh.
\newblock A poisson process for submodular maximization.
\newblock In \emph{Proceedings of the 58th Annual {ACM} Symposium on Theory of
  Computing}, pages 2152--2163, 2026.

\bibitem[Gauthier(2021)]{gauthier2021glassdoor}
David Gauthier.
\newblock Glassdoor job reviews.
\newblock Kaggle dataset, 2021.
\newblock
  \url{https://www.kaggle.com/datasets/davidgauthier/glassdoor-job-reviews},
  accessed August 2026.

\bibitem[Golrezaei et~al.(2014)Golrezaei, Nazerzadeh, and
  Rusmevichientong]{GolrezaeiNR14}
Negin Golrezaei, Hamid Nazerzadeh, and Paat Rusmevichientong.
\newblock Real-time optimization of personalized assortments.
\newblock \emph{Management Science}, 60\penalty0 (6):\penalty0 1532--1551,
  2014.

\bibitem[Gong et~al.(2022)Gong, Goyal, Iyengar, Simchi-Levi, Udwani, and
  Wang]{GongGISUW22}
Xiao-Yue Gong, Vineet Goyal, Garud~N. Iyengar, David Simchi-Levi, Rajan Udwani,
  and Shuangyu Wang.
\newblock Online assortment optimization with reusable resources.
\newblock \emph{Management Science}, 68\penalty0 (7):\penalty0 4772--4785,
  2022.

\bibitem[Kon(2024)]{kon2024linkedin}
Arsh Kon.
\newblock {LinkedIn} job postings (2023--2024).
\newblock Kaggle dataset, 2024.
\newblock \url{https://www.kaggle.com/datasets/arshkon/linkedin-job-postings},
  accessed August 2026.

\bibitem[Li et~al.(2015)Li, Rusmevichientong, and Topaloglu]{LiRT15}
Guang Li, Paat Rusmevichientong, and Huseyin Topaloglu.
\newblock The d-level nested logit model: Assortment and price optimization
  problems.
\newblock \emph{Operations Research}, 63\penalty0 (2):\penalty0 325--342, 2015.

\bibitem[Ma and Simchi-Levi(2020)]{MaS20}
Will Ma and David Simchi-Levi.
\newblock Algorithms for online matching, assortment, and pricing with tight
  weight-dependent competitive ratios.
\newblock \emph{Operations Research}, 68\penalty0 (6):\penalty0 1787--1803,
  2020.

\bibitem[Nemhauser et~al.(1978)Nemhauser, Wolsey, and Fisher]{NemhauserWF78}
George~L. Nemhauser, Laurence~A. Wolsey, and Marshall~L. Fisher.
\newblock An analysis of approximations for maximizing submodular set
  functions—{I}.
\newblock \emph{Mathematical Programming}, 14:\penalty0 265--294, 1978.

\bibitem[Pollner et~al.(2022)Pollner, Roghani, Saberi, and Wajc]{PollnerRSW22}
Tristan Pollner, Mohammad Roghani, Amin Saberi, and David Wajc.
\newblock Improved online contention resolution for matchings and applications
  to the gig economy.
\newblock In \emph{Proceedings of the 23rd ACM Conference on Economics and
  Computation}, pages 321--322, 2022.

\bibitem[Rios and Torrico(2023)]{RiosT23}
Ignacio Rios and Alfredo Torrico.
\newblock Platform design in matching markets: A two-sided assortment
  optimization approach, 2023.
\newblock Working paper. Available online as arXiv report 2308.02584.

\bibitem[Rios et~al.(2023)Rios, Saban, and Zheng]{RiosSZ23}
Ignacio Rios, Daniela Saban, and Fanyin Zheng.
\newblock Improving match rates in dating markets through assortment
  optimization.
\newblock \emph{Manufacturing \& Service Operations Management}, 25\penalty0
  (4):\penalty0 1304--1323, 2023.

\bibitem[Rusmevichientong et~al.(2010)Rusmevichientong, Shen, and
  Shmoys]{RusmevichientongSS10}
Paat Rusmevichientong, Zuo-Jun~Max Shen, and David~B. Shmoys.
\newblock Dynamic assortment optimization with a multinomial logit choice model
  and capacity constraint.
\newblock \emph{Operations Research}, 58\penalty0 (6):\penalty0 1666--1680,
  2010.

\bibitem[Rusmevichientong et~al.(2014)Rusmevichientong, Shmoys, Tong, and
  Topaloglu]{RusmevichientongSTT14}
Paat Rusmevichientong, David Shmoys, Chaoxu Tong, and Huseyin Topaloglu.
\newblock Assortment optimization under the multinomial logit model with random
  choice parameters.
\newblock \emph{Production and Operations Management}, 23\penalty0
  (11):\penalty0 2023--2039, 2014.

\bibitem[Sumida et~al.(2021)Sumida, Gallego, Rusmevichientong, Topaloglu, and
  Davis]{SumidaGRTD21}
Mika Sumida, Guillermo Gallego, Paat Rusmevichientong, Huseyin Topaloglu, and
  James Davis.
\newblock Revenue-utility tradeoff in assortment optimization under the
  multinomial logit model with totally unimodular constraints.
\newblock \emph{Management Science}, 67\penalty0 (5):\penalty0 2845--2869,
  2021.

\bibitem[Talluri and Van~Ryzin(2004)]{Talluri04}
Kalyan Talluri and Garrett Van~Ryzin.
\newblock Revenue management under a general discrete choice model of consumer
  behavior.
\newblock \emph{Management Science}, 50\penalty0 (1):\penalty0 15--33, 2004.

\bibitem[Topaloglu(2013)]{Topaloglu13}
Huseyin Topaloglu.
\newblock Joint stocking and product offer decisions under the multinomial
  logit model.
\newblock \emph{Production and Operations Management}, 22\penalty0
  (5):\penalty0 1182--1199, 2013.

\bibitem[Torrico et~al.(2026)Torrico, Carvalho, and Lodi]{TorricoCL22}
Alfredo Torrico, Margarida Carvalho, and Andrea Lodi.
\newblock Multi-agent assortment optimization in sequential matching markets,
  2026.
\newblock {\em Mathematics of Operations Research}, forthcoming.

\bibitem[Udwani(2025)]{Udwani25}
Rajan Udwani.
\newblock Submodular order functions and assortment optimization.
\newblock \emph{Management Science}, 71\penalty0 (1):\penalty0 202--218, 2025.

\end{thebibliography}

\changelocaltocdepth{1} 
\appendix
\section{Additional Proofs from Section~\ref{sec:purely-combinatorial}}
\subsection{Proof of Claim~\ref{clm:muij_vs_pij}} \label{app:proof_clm_muij_vs_pij}
Let us begin by observing that
\begin{eqnarray}
\mu_{ij} & = & \ex{ \frac{w_{ij} I^{ \hat{M} }_{ij}}{1 + \sum_{\ell \in \tilde{T}_j} w_{\ell j} I^{ \hat{M} }_{\ell j}} } \nonumber \\
& \geq & w_{ij} \cdot \pi_i(j, \tilde{M}_i) \cdot \ex{ \frac{1}{1 + w_{ij} + \sum_{\ell \in \tilde{T}_j \setminus \{ i \}} w_{\ell j} I^{ \hat{M} }_{\ell j}} } \label{eqn:clm_muij_vs_pij_eq1} \\
& = & p_{ij} \cdot \ex{ \frac{ 1 + w_{ij} }{1 + w_{ij} + \sum_{\ell \in \tilde{T}_j \setminus \{ i \}} w_{\ell j} I^{ \hat{M} }_{\ell j}} } \ . \label{eqn:clm_muij_vs_pij_eq2}
\end{eqnarray}
Here, inequality~\eqref{eqn:clm_muij_vs_pij_eq1} holds since the indicators $\{ I^{ \hat{M} }_{\ell j} \}_{\ell \in \C}$ are mutually independent, with $I^{ \hat{M} }_{\ell j} \sim \mybern( {\pi_{\ell}}(j, \hat{M}_{\ell}) )$. In addition, since $\hat{M}_i \subseteq \tilde{M}_i$, we have $\pi_i(j, \hat{M}_i) \geq \pi_i(j, \tilde{M}_i)$. Equality~\eqref{eqn:clm_muij_vs_pij_eq2} is obtained by recalling that $p_{ij} = \hat{w}_{ij} \cdot \pi_i(j, \tilde{M}_i)$ and $\hat{w}_{ij} = \frac{ w_{ij} }{ 1 + w_{ij} }$. Now, for the purpose of showing that $\mu_{ij} \geq p_{ij} \mu_{0j}$, we argue that the expectation in equation~\eqref{eqn:clm_muij_vs_pij_eq2} forms an upper bound on $\mu_{0j}$. Indeed, 
\[ \mu_{0j} ~~=~~ \ex{ \frac{1}{1 + \sum_{\ell \in \tilde{T}_j} w_{\ell j} I^{ \hat{M} }_{\ell j}} } ~~\leq~~ \ex{ \frac{ 1 }{1 + \sum_{\ell \in \tilde{T}_j \setminus \{ i \}} w_{\ell j} I^{ \hat{M} }_{\ell j}} } ~~\leq~~ \ex{ \frac{ 1 + w_{ij} }{1 + w_{ij} + \sum_{\ell \in \tilde{T}_j \setminus \{ i \}} w_{\ell j} I^{ \hat{M} }_{\ell j}} } \ . \]
\section{Additional Proofs from Section~\ref{sec:preliminaries}}

\subsection{Proof of Lemma~\ref{lem:assortment_distribution_is_well_defined}}\label{app:proof_assortment_distribution_is_well_defined}

To show that the constants $\psi_0, \ldots, \psi_n$, previously referred to as ``probabilities'', indeed sum-up to $1$, note that
\begin{eqnarray*}
    \sum_{j = 0}^n \psi_j & = & \sum_{j = 0}^{n-1} \left(\frac{x_j}{u_j} - \frac{x_{j+1}}{u_{j+1}}\right) \cdot \sum_{\ell = 0}^j u_j + \frac{x_n}{u_n} \cdot \sum_{\ell = 0}^n u_\ell \nonumber \\
    & = & \sum_{j \in [n]} \frac{x_j}{u_j} \cdot \left( \sum_{\ell = 0}^j u_\ell - \sum_{\ell = 0}^{j - 1} u_\ell \right) + x_0 \\
    & = & \sum_{j \in [n]} x_j + x_0 \nonumber \\
    & = & 1 \ ,
\end{eqnarray*}
where the last equality is obtained by recalling that $x_0 = 1 - \sum_{j \in [n]} x_j$.

\subsection{Proof of Lemma~\ref{lem:assortment_distribution_realizes_LP_solution}}\label{app:proof_assortment_distribution_realizes_LP_solution}

To evaluate the expected choice probability $\exsubpar{A \sim \mathcal{D}(x)}{\pi (j, A)}$ of each alternative $j \in [n]$, we condition on the random assortment sampled from the distribution $\mathcal{D}(x)$. 
To this end, we observe that for all $k < j$, when $A_k = [k]$ is sampled, alternative $j$ obviously cannot be chosen, i.e., $\pi(j, A_k) = 0$. 
For $k \geq j$, the conditional probability of selecting this alternative is $\pi(j, A_k)= \frac{ u_j }{ \sum_{\ell = 0}^k u_\ell }$. 
Therefore,
\begin{eqnarray}
    \exsub{A \sim \mathcal{D}(x)}{\pi (j, A)} & = & \sum_{k = j}^n \prsub{A \sim \mathcal{D}(x)}{ A = A_k } \cdot \pi(j, A_k) \nonumber \\
    & = & \sum_{k = j}^{n - 1} \frac{u_j}{\sum_{\ell = 0}^k u_\ell} \cdot \left( \frac{x_k}{u_k} - \frac{x_{k+1}}{u_{k + 1}} \right) \cdot \sum_{\ell = 0}^k u_\ell + \frac{u_j}{\sum_{\ell = 0}^n u_\ell} \cdot \frac{x_n}{u_n} \cdot \sum_{\ell = 0}^n u_\ell \\
    & = & u_j \cdot \left( \sum_{k = j}^{n - 1} \left( \frac{x_k}{u_k} - \frac{x_{k+1}}{u_{k + 1}} \right) + \frac{x_n}{u_n} \right) \nonumber \\
    & = & x_j \ . \nonumber
\end{eqnarray}
Here, the second equality is obtained by substituting $\prpar{A = A_k} = \psi_k = ( \frac{x_k}{u_k} - \frac{x_{k+1}}{u_{k + 1}} ) \cdot \sum_{\ell = 0}^k u_\ell$ for $k \leq n-1$, along with $\prpar{A = A_n} = \psi_n = \frac{x_n}{u_n} \cdot \sum_{\ell = 0}^n u_\ell$.

\subsection{Proof of Lemma~\ref{lem:randomized_assortment}} \label{app:proof_lem_randomized_assortment}

To establish the desired claim, note that
\begin{eqnarray}
\rev(\mathcal{D}(x)) & = & \exsub{M \sim \mathcal{D}(x)}{\rev(M)} \nonumber \\
& = & \sum_{j \in \S} \exsub{M \sim \mathcal{D}(x)}{ \ex{  f_j(C^M_j) } } \label{equality:rev_M_definition} \\
& = & \sum_{j \in \S} \sum_{C_j \subseteq \C} {\prsub{M \sim \mathcal{D}(x)}{ C_j^M = C_j } \cdot f_j(C_j) }\label{equality:expectation_definition} \\
& = & \sum_{j \in \S} \sum_{C_j \subseteq \C} {\pr{ C_j^x = C_j }} \cdot f_j(C_j) \label{equality:menus_to_points} \\
& = & \sum_{j \in \S} \exsubpar{x}{f_j(C^x_j)} \nonumber \\
& = & \rev^P(x) \ . \nonumber
\end{eqnarray}
Here, equality~\eqref{equality:rev_M_definition} is obtained by substituting $\rev(M) = \sum_{j \in \S} \expar{f_j(C^M_j)}$.
To clarify equation~\eqref{equality:expectation_definition}, we note that $\prparsub{M \sim \mathcal{D}(x)}{ C_j^M = C_j }$ contains two layers of randomness: Sampling a menu $M$ according to the distribution $\mathcal{D}(x)$, and subsequently sampling the set of customers $C^M_j$ who select supplier $j$.
To derive equality~\eqref{equality:menus_to_points}, it remains to explain why $\prparsub{M \sim \mathcal{D}(x)}{ C_j^M = C_j } = \prpar{ C_j^x = C_j }$.
For this purpose, we observe that
\begin{eqnarray}
\prsub{M \sim \mathcal{D}(x)}{ C_j^M = C_j } & = & \prsub{M \sim \mathcal{D}(x)}{ \left( \bigwedge_{i \in C_j} \{ I^M_{ij} = 1 \} \right) \wedge \left( \bigwedge_{i \in \C \setminus C_j} \{ I^M_{ij} = 0 \} \right) } \nonumber \\
& = & \exsub{M \sim \mathcal{D}(x) }{ \ex{ \prod_{i \in C_j} I^M_{ij} \cdot \prod_{i \in \C \setminus C_j} ( 1 - I^M_{ij} ) } } \nonumber \\
& = & \prod_{i \in C_j} \exsub{M_i \sim \mathcal{D}(x_{i \cdot}) }{ \pi_i(j, M_i) } \nonumber \\
&& \qquad \cdot \prod_{i \in \C \setminus C_j} \exsub{M_i \sim \mathcal{D}(x_{i \cdot}) }{ 1-\pi_i(j, M_i) } \label{equality:I_are_independent} \\
& = & \prod_{i \in C_j} x_{ij} \cdot \prod_{i \in \C \setminus C_j} ( 1-x_{ij} ) \label{equality:random_assortment_property} \\
& = & \pr{ \left( \bigwedge_{i \in C_j} \{ I^x_{ij} = 1 \} \right) \wedge \left( \bigwedge_{i \in \C \setminus C_j} \{ I^x_{ij} = 0 \} \right) } \nonumber \\
& = & \pr{ C_j^x = C_j } \ . \nonumber
\end{eqnarray}
Equality~\eqref{equality:I_are_independent} holds since $(I^M_{ij})_{i \in \C}$ are mutually independent, and since each $I^M_{ij}$ indicates whether customer $i$ selects supplier $j$ with respect to her menu $M_i$, which happens with probability $\pi_i(j, M_i)$, as explained in Section~\ref{subsec:description_of_the_problem}.
Equality~\eqref{equality:random_assortment_property} follows from Lemma~\ref{lem:assortment_distribution_realizes_LP_solution}, stating in this case that $x_{ij}$ identifies with the probability of customer $i$ choosing supplier $j$ from a random menu sampled according to the distribution $\mathcal{D}(x_{i \cdot})$.
\section{Additional Proofs from Section~\ref{sec:customized}}\label{app:main_algorithm}

\subsection{Proof of Lemma~\ref{lem:auxiliary_properties}} \label{app:auxiliary_properties}

\paragraph{Property 1: ${\hat{y} \in P^\S}$.} By definition, we have  $\hat{y}_{\cdot j} = \sum_{C_j \subseteq \C} \prpar{C^{x^*}_j = C_j} \cdot y^{*}_{\cdot j}(C_j)$ for every supplier $j \in \S$, meaning that $\hat{y}_{\cdot j}$ is a convex combination of the vectors $\{ y^{*}_{\cdot j}(C_j) \}_{C_j \subseteq \C}$. Additionally, $y^{*}_{\cdot j}(C_j)$ is feasible in~\eqref{eqn:assortment_optimization}, for all $C_j\subseteq\C$. Therefore, by noting that the feasible region of~\eqref{eqn:assortment_optimization} is contained in $P^\S_j$, since the latter is a downward-closed polyhedron, it follows that $\hat{y}_{\cdot j} \in P^\S_j$ for every $j \in \S$, implying that $\hat{y} \in P^\S$.

\paragraph{Property 2: ${\hat{y}_{ij} \leq \hat{w}_{ij} x^*_{ij}}$ for every ${(i, j) \in \E}$.} To establish this relation, we observe that
\begin{eqnarray}
    \hat{y}_{ij} & = & \sum_{C_j \subseteq \C} \pr{C^{x^*}_j = C_j} \cdot y^{*}_{ij}(C_j) \nonumber \\
    & = & \sum_{\MyAbove{C_j \subseteq \C:}{i \in C_j}} \pr{C^{x^*}_j = C_j} \cdot y^{*}_{ij}(C_j) \label{equality:y_star_zero} \\
    & \leq & \hat{w}_{ij} \cdot \sum_{\MyAbove{C_j \subseteq \C:}{i \in C_j}} \pr{C^{x^*}_j = C_j} \label{ieq:y_in_P^{C_j}} \\
    & = & \hat{w}_{ij} x^*_{ij} \ . \label{equality:mutual_exclusive_probability}
\end{eqnarray}
Here, equality~\eqref{equality:y_star_zero} holds since $y^{*}_{ij}(C_j) = 0$ when $i \notin C_j$, by recalling that $y^{*}_{ij}(C_j)$ is feasible to~\eqref{eqn:assortment_optimization}.
Inequality~\eqref{ieq:y_in_P^{C_j}} is obtained by noting that $y^{*}_{ij}(C_j)\in P^\S_j$ forces $\frac{y^{*}_{ij}(C_j)}{w_{ij}} + \sum_{\ell \in \C} y^{*}_{\ell j}(C_j) \leq 1$, for every customer $i \in \C$.
In particular, we must have $\frac{y^{*}_{ij}(C_j)}{w_{ij}} + y^{*}_{ij}(C_j) \leq 1$, implying that $y^{*}_{ij}(C_j) \leq \frac{w_{ij}}{1 + w_{ij}} = \hat{w}_{ij}$.   
To elaborate on equality~\eqref{equality:mutual_exclusive_probability}, by recalling that $C^{x^*}_j = \{ i \in \C : I^{x^*}_{ij} = 1 \}$ and $I^{x^*}_{ij} \sim \mybern(x^*_{ij})$, we have  
\[ \sum_{\MyAbove{C_j \subseteq \C:}{i \in C_j}} \pr{C^{x^*}_j = C_j} ~~=~~ \pr{i \in C^{x^*}_j} ~~=~~ \pr{I^{x^*}_{ij} = 1} ~~=~~ x^*_{ij} \ . \]

\subsection{Proof of Claim~\ref{clm:mustar_vs_pstar}} \label{app:proof_mustar_vs_pstar}
The proof is along the same lines as that of  Claim~\ref{clm:muij_vs_pij}. We start by noting that
\begin{eqnarray}
\mu^*_{ij} & = & \ex{ \frac{w_{ij} I^{ x^* }_{ij}}{1 + \sum_{\ell \in A_j} w_{\ell j} I^{ x^* }_{\ell j}} } \nonumber \\
& = & w_{ij} \cdot x_{ij}^* \cdot \ex{ \frac{1}{1 + w_{ij} + \sum_{\ell \in A_j \setminus \{ i \}} w_{\ell j} I^{ x^*}_{\ell j}} } \label{eqn:clm_mustar_vs_pstar_eq1} \\
& = & p^*_{ij} \cdot \ex{ \frac{ 1 + w_{ij} }{1 + w_{ij} + \sum_{\ell \in A_j \setminus \{ i \}} w_{\ell j} I^{ x^*}_{\ell j}} } \ . \label{eqn:clm_mustar_vs_pstar_eq2}
\end{eqnarray}
Equality~\eqref{eqn:clm_mustar_vs_pstar_eq1} follows since the indicators $\{ I^{ x^*}_{\ell j} \}_{\ell \in \C}$ are mutually independent, with $I^{ x^* }_{\ell j} \sim \mybern( x^*_{\ell j} )$. Then, \eqref{eqn:clm_mustar_vs_pstar_eq2} holds by recalling that  $p^*_{ij} = \hat{w}_{ij} \cdot x^*_{ij}$ and $\hat{w}_{ij} = \frac{ w_{ij} }{ 1 + w_{ij} }$. To show that $\mu^*_{ij} \geq p^*_{ij} \mu^*_{0j}$, we argue that the expectation in equation~\eqref{eqn:clm_mustar_vs_pstar_eq2} forms an upper bound on $\mu^*_{0j}$. Indeed, 
\[ \mu^*_{0j} ~~=~~ \ex{ \frac{1}{1 + \sum_{\ell \in A_j} w_{\ell j} I^{ x^* }_{\ell j}} } ~~\leq~~ \ex{ \frac{ 1 }{1 + \sum_{\ell \in A_j \setminus \{ i \}} w_{\ell j} I^{ x^* }_{\ell j}} } ~~\leq~~ \ex{ \frac{ 1 + w_{ij} }{1 + w_{ij} + \sum_{\ell \in A_j \setminus \{ i \}} w_{\ell j} I^{ x^* }_{\ell j}} } \ . \]

\subsection{Proof of Lemma~\ref{lem:lowerbound_onehalf}} \label{app:lowerbound_onehalf}

We start by describing the instance:
\paragraph{Instance.} Consider a single supplier $\S = \{j\}$ and a set of identical customers $\C=\{1,\ldots,n\}$ where 
$n=s^{3}$. Each pair $(i,j)\in\E$ has a reward $r_{ij} =1$ and preference weights $w:=w_{ij}=s^{-2}$, $u_{ij}=\frac{p}{1-p}$ with $p:=s^{-1}$.
Observe that $p=\frac{u_{ij}}{1+u_{ij}}$; hence, by offering menu
$M_i=\{j\}$ to customer $i$ we have that $i$ chooses $j$ with probability
exactly $x_{ij}=p$, which makes the constraint $\frac{x_{ij}}{u_{ij}}+x_{ij}\le1$ binding in $P^\C$. Since $n=s^3$, the following identities hold: 
$np=s^{2},\
npw=1,\ nw=s$.

\paragraph{Computing $\opt\eqref{eqn:customized_lp}$.}
In the following, we show that $\opt\eqref{eqn:customized_lp} = \dfrac{s}{s+1}$.
First, since all rewards are 1, the objective function in \eqref{eqn:customized_lp} is $Y:=\sum_{i\in\C}y_{ij}$ so it suffices to upper bound $Y$ over
$P^{\mathcal S}$. Each constraint $\frac{y_{ij}}{w}+\sum_{i\in\C}y_{ij}\leq 1$ is equivalent to $y_{ij}\le w(1-Y)$. Thus, by summing over all customers we get $Y\le n\,w\,(1-Y)$ which is equivalent to $Y\le\frac{nw}{1+nw}=\frac{s}{s+1}$. Second, this bound is attained at the feasible point $y_{ij}=\frac{w}{1+nw}$ for all $i\in\C$,
which indeed lies in $P^{\mathcal S}$ with equality in every non-trivial constraint. On the other hand, the point $x_{ij}=p$ for all $i\in\C$ is feasible in $P^\C$. Now, let us check that $y_{ij}\leq \hat{w}\,x_{ij}$. For this, observe that 
\[
\frac{w}{1+nw}\ \le\ \frac{w}{1+w} = \hat{w}\cdot p\;,
\]
which is equivalent to $1+w\leq p+pnw$.
Since $npw=1$, the right-hand side equals $p+1$, and the
inequality becomes $w\le p$, which clearly holds since $w<p$. Therefore, by decreasing $x_{ij}$ to get $\bar{x}_{ij}$ such that $y_{ij}=\hat{w}\cdot \bar x_{ij}$, we obtain a feasible solution in \eqref{eqn:customized_lp} with objective value $ \frac{s}{1+s}$. In other words, $\opt\eqref{eqn:customized_lp} = \frac{s}{1+s}$.

\paragraph{Computing $\optcust\eqref{eqn:problem_R}$.} In the following, we show that $\revcustP(x)\in [\frac{1}{2}-\frac{1}{s},\frac{1}{2}]$ where $x_{ij}=p$ for every customer $i\in\C$. 
Let $|C_j|\sim\mathrm{Bin}(n,p)$ be the random number of customers who select $j$. Since supplier $j$ has the same value $w$ across all customers and the reward values are all 1,
the supplier's optimal customized assortment is the whole backlog:
\[
\fcustomized_j(C_j)=\frac{w\cdot|C_j|}{1+w\cdot|C_j|}\;.
\]
Define $\varphi(z) = z/(1+z)$. Hence, due to the  monotonicity of $\varphi$, showing $j$ to all customers is optimal and $\optcust\eqref{eqn:problem_R}=\revcustP(x) = \ex{\varphi(w\cdot|C_j|)}$ where $x_{ij}=p$ for all $i\in\C$. Here we used that $x_{ij}$ is coordinate-wise maximal in $\frac{x_{ij}}{u_{ij}}+x_{ij} \leq 1$ and $\fcustomized_j$ is monotone, so there is no other solution with a higher objective value.
Now notice that $\ex{w\cdot|C_j|}=npw=1$ and
\[
\text{Var}[w\cdot|C_j|]=w^2\,np(1-p)\ \le\ w^2\,np = w = s^{-2}.
\]
Since $|\varphi'(z)|=
\frac1{(1+z)^2}\le1$, $\varphi$ is $1$-Lipschitz and
\[
\big| \ex{\varphi(w\cdot|C_j|)}-\varphi(1)\big|
\ \le\ \ex{\big|w\cdot|C_j|-1\big|}
\ \le\ \sqrt{\text{Var}[w\cdot|C_j|]}
\ \le\ \frac{1}{s}.
\]
Moreover, thanks to Jensen's inequality,
$$
\ex{\varphi(w\cdot|C_j|)}\ \le\ \varphi\big(\ex{w\cdot |C_j|}\big)=\varphi(1)=\tfrac12.
$$
Combining the two bounds we get $\frac{1}{2}-\tfrac1s\le\mathbb E[\varphi(w\cdot|C_j|)]\le\frac12$\;. Therefore, as $s\to+\infty$ we obtain 
\[
\frac{\revcustP(x)}{\opt\eqref{eqn:customized_lp}} = \frac{\optcust\eqref{eqn:problem_R}}{\opt\eqref{eqn:customized_lp}} \to \frac12.
\]

\subsection{Proof of Theorem~\ref{thm:revenue_gap}}\label{app:rewardgap} 
Thanks to Theorem~\ref{thm:main_algorithm}, there is a point $\bar x\in P^\C$ such that
\(
\frac{\revincP(\bar x)}{\optcust\eqref{eqn:problem_R}}\geq \frac{1}{2},
\)
thus \(\frac{\optcust\eqref{eqn:problem_R}}{\optinc\eqref{eqn:problem_R}}\leq 2\), i.e., the reward gap is upper bounded by 2. Now, we present an instance where the gap is arbitrarily close to $2$.

\paragraph{Instance.} Let $\varepsilon \in (0,1)$ and $h=(1+\varepsilon)/(2+\varepsilon)$. Consider a single supplier $\S = \{j\}$ and two customers $\mathcal C = \{1,2\}$. 
Set the pairwise rewards and the preference weights of $j$ as follows: $r_{1j}=1, w_{1j}= 1+\varepsilon, r_{2j} = \varepsilon, w_{2j}= \tfrac{1}{\varepsilon}$.
Moreover, choose the customer weights $u_{1j}, u_{2j}$ so that
\[
\pi_1^{(c)}(j, \{j\}) = \frac{u_{1j}}{1+u_{1j}} = a, \qquad \pi_2^{(c)}(j, \{j\}) = \frac{u_{2j}}{1+u_{2j}} = b,
\]
where $a := \frac{\varepsilon(1-\varepsilon)(2+\varepsilon)}{(1+\varepsilon)^2}$ and $b:=1-\varepsilon$. 

We first observe that, since there is only one supplier and two customers, we can evaluate $f_j^{\mathrm{cust}}$ and $f_j^{\mathrm{inc}}$ on all four possible realizations of the random set of customers $C_j \subseteq \{1,2\}$ that choose supplier $j$ over the outside option. For either $f_j=\fcustomized_j$ or $f_j=\finclusive_j$ we have:
\[
f_j(\emptyset) = 0, \qquad f_j(\{1\}) = \frac{r_{1j}w_{1j}}{1+w_{1j}}= \frac{1+\varepsilon}{2+\varepsilon}, \qquad f_j(\{2\}) = \frac{r_{2j}w_{2j}}{1+w_{2j}}= \frac{\varepsilon}{1+\varepsilon}.
\]
For $C_j = \{1,2\}$, we obtain the following 
\begin{align*}
\finclusive_j(\{1,2\}) &= \frac{r_{1j}w_{1j}+r_{2j}w_{2j}}{1+w_{1j}+w_{2j}} = \frac{2+\varepsilon}{2+\varepsilon+1/\varepsilon} \\
\fcustomized_j(\{1,2\}) &=\max\Big\{\frac{r_{1j}w_{1j}}{1+w_{1j}},\, \frac{r_{2j}w_{2j}}{1+w_{2j}},\, \finclusive_j(\{1,2\})\Big\} = \max\Big\{\frac{1+\epsilon}{2+\epsilon},\, \frac{\epsilon}{1+\epsilon},\, \finclusive_j(\{1,2\})\Big\} \geq \frac{1+\epsilon}{2+\epsilon}.
\end{align*}

We now show that on this instance \(\frac{\optcust\eqref{reformulation1}}{\optinc\eqref{reformulation1}}\) tends to 2. To prove this, we just need a feasible solution for the customized model and an optimal solution for the inclusive model.
\paragraph{Lower bounding the customized model.} We consider the feasible solution in which both customers see $j$. Given this, we have all four possible realizations $C_j$ as above and the corresponding probabilities are
\begin{align*}
\Pr[C_j = \{1,2\}] = ab,& \quad \Pr[C_j = \{1\}] = a(1-b),\\
\Pr[C_j = \{2\}] = (1-a)b,& \quad \Pr[C_j = \emptyset] = (1-a)(1-b).
\end{align*}
Let $M=(M_1,M_2) = (\{j\},\{j\})$. Therefore,
\begin{align*}
\opt^{\text{cust}}&\geq \rev^{\text{cust}}(M)= a(1-b)\frac{1+\varepsilon}{2+\varepsilon}+ (1-a)b\frac{\varepsilon}{1+\varepsilon} + ab\, f_j^{\mathrm{cust}}(\{1,2\}) \\
&\;\ge\; a\varepsilon\frac{1+\varepsilon}{2+\varepsilon}+(1-a)(1-\varepsilon)\frac{\varepsilon}{1+\varepsilon}+a(1-\varepsilon)\frac{1+\varepsilon}{2+\varepsilon}\\
&=a\varepsilon\frac{1+\varepsilon}{2+\varepsilon}+(1-a)a\frac{1+\varepsilon}{2+\varepsilon}+a(1-\varepsilon)\frac{1+\varepsilon}{2+\varepsilon}\\
&=a\frac{1+\varepsilon}{2+\varepsilon}(2-a).
\end{align*}
where in the second inequality we replaced $b=1-\varepsilon$ and lower bounded $f_j^{\text{cust}}(\{1,2\})$ by $\frac{1+\varepsilon}{2+\varepsilon}$. In the second equality we used the definition of $a$ to replace the second term by $(1-a)(1-\varepsilon)\frac{\varepsilon}{1+\varepsilon}= (1-a)a\frac{1+\varepsilon}{2+\varepsilon} $.

\paragraph{Optimal value under the inclusive model.} We evaluate all possible menus.
\begin{align*}
    \rev^{\mathrm{inc}}(\emptyset,\{j\}) &= b\frac{\varepsilon}{1+\varepsilon} = (1-\varepsilon)\frac{\varepsilon}{1+\varepsilon} = a\frac{1+\varepsilon}{2+\varepsilon}\\
    \rev_{\mathrm{inc}}(\{j\},\emptyset) &= a\frac{1+\varepsilon}{2+\varepsilon},
\end{align*}
and in the last case we obtain
\begin{align}
\rev^{\mathrm{inc}}(\{j\},\{j\}) &= a(1-b)\frac{1+\varepsilon}{2+\varepsilon} + (1-a)b\frac{\varepsilon}{1+\varepsilon} + ab \cdot \frac{2+\varepsilon}{2+\varepsilon+1/\varepsilon}\notag \\
&= a\varepsilon\frac{1+\varepsilon}{2+\varepsilon} + (1-a)(1-\varepsilon)\frac{\varepsilon}{1+\varepsilon} + a(1-\varepsilon) \cdot \frac{2+\varepsilon}{2+\varepsilon+1/\varepsilon}\label{eqn:aux_lb_gap1}\\
&= a\varepsilon\frac{1+\varepsilon}{2+\varepsilon} + (1-a)a\frac{1+\varepsilon}{2+\varepsilon} + a(1-\varepsilon) \cdot \frac{2+\varepsilon}{2+\varepsilon+1/\varepsilon}\label{eqn:aux_lb_gap2}\\
& = a\frac{1+\varepsilon}{2+\varepsilon}\left[\varepsilon+(1-a)+(1-\varepsilon) \cdot \frac{(2+\varepsilon)^2}{(1+\varepsilon)(2+\varepsilon+1/\varepsilon)}\right] \notag\\
& = a\frac{1+\varepsilon}{2+\varepsilon}\left[\varepsilon+1-(1-\varepsilon)\cdot\frac{\varepsilon(2+\varepsilon)}{(1+\varepsilon)^2}+(1-\varepsilon) \cdot \frac{\varepsilon(2+\varepsilon)^2}{(1+\varepsilon)^3}\right]\label{eqn:aux_lb_gap3}\\
& = a\frac{1+\varepsilon}{2+\varepsilon}\left[\varepsilon+1+(1-\varepsilon)\cdot\frac{\varepsilon(2+\varepsilon)}{(1+\varepsilon)^3}\right]\notag\\
&=a\frac{1+\varepsilon}{2+\varepsilon}L(\varepsilon)\notag
\end{align}
where
\[
L(\varepsilon) = \varepsilon+1+(1-\varepsilon)\cdot\frac{\varepsilon(2+\varepsilon)}{(1+\varepsilon)^3}.
\]
In \eqref{eqn:aux_lb_gap1}, we replace $b=1-\varepsilon$. In \eqref{eqn:aux_lb_gap2}, we use $a\frac{1+\varepsilon}{2+\varepsilon} = (1-\varepsilon)\frac{\varepsilon}{1+\varepsilon}$. In \eqref{eqn:aux_lb_gap3}, we replace $a=(1-\varepsilon)\frac{\varepsilon(2+\varepsilon)}{(1+\varepsilon)^2}$ in the second term inside the parentheses. Finally, note that $L(\varepsilon)> 1$ when $\varepsilon>0$, which means $M=(\{j\},\{j\})$ is the optimal menu.
We conclude by noting that
\[
\frac{\optcust\eqref{reformulation1}}{\optinc\eqref{reformulation1}} \;\ge\; \frac{a\frac{1+\varepsilon}{2+\varepsilon}(2-a)}{a\frac{1+\varepsilon}{2+\varepsilon}L(\varepsilon)} = \frac{2-a}{L(\varepsilon)}
\]
where the right-hand side tends to $2$ because $L(\varepsilon)\to 1$ and $a\to 0$ when $\varepsilon\to0$.

{\color{black}
\section{Missing Details in  Section~\ref{sec:computational_study}}

\subsection{Synthetic Setup}\label{app:syntheticdetails}
Let us denote Unif$(a,b)$ the uniform random variable in the interval $[a,b]$ and Exp($\lambda$) the exponential random variable with parameter $\lambda$. In the following, we describe how each instance was generated:
\begin{itemize}
    \item {\it Heterogeneous-dense.} First, every pair customer $i$ and supplier $j$ is compatible. Parameters $u_{i,j}$ were drawn according to $\mathrm{Exp}(1)$. On the other hand, parameters $w_{i,j}$ were randomly split in two groups: popular and not popular. Specifically, a customer $i$ is popular with probability 0.3 and $w_{i,j}\sim \mathrm{Unif}[2,3]$ with reward $r_{i,j}\sim \mathrm{Unif}[0,1]$. Otherwise, a supplier is unpopular and $w_{i,j}\sim \mathrm{Unif}[0,0.1]$ with a higher reward $r_{i,j}\sim \mathrm{Unif}[1,10]$. The weights for the outside options were set to 1.
    \item {\it Homogeneous-sparse.} First, every pair customer $i$ and supplier $j$ is compatible with probability $0.5$. Parameters $u_{i,j}=u_j$ were drawn according to $\mathrm{Exp}(1)$. We set the  supplier to be less profitable with probability 0.3 and $r_{i,j}\sim \mathrm{Unif}[0,1]$. If supplier is profitable, then $r_{i,j}\sim \mathrm{Unif}[1,10]$.  On the other hand, parameters $w_{i,j}=w_j$ were randomly split in two groups: picky and not picky. Specifically, with probability 0.3 we draw $w_{j}\sim \mathrm{Unif}[2,3]$ (not picky since $w_{0,j} = 1/w_j$), otherwise the supplier is picky and $w_{j}\sim \mathrm{Unif}[0,0.1]$ (i.e., the outside option is heavier).
\end{itemize}

\subsection{Semi-Synthetic Setup}\label{app:semidetails}

\textbf{Data treatment.} Postings in the LinkedIn corpus are filtered to data-science titles,
deduplicated by posting identifier, and postings with missing (as opposed to
zero) application counts are excluded from estimation while retained for
attributes. A supplier is a (company, seniority level) pair, whose attributes
aggregate over its postings:
\begin{itemize}
\item the salary $\sal_j$, in thousands of USD, is the median over the pair's
postings of the posting-level \emph{annualized} salary. Posted pay was
converted to an annual figure using its pay period (hourly, monthly, yearly).
Postings that report no pay are completed with the salary survey's median salary
for the same seniority level, adjusted by a remote-work multiplier;
\item the employer rating $\ratg_j \in [0,1]$ is the firm's Glassdoor rating
(rescaled from the 1--5 scale) when the firm can be matched by name. Otherwise, we use a prestige proxy built from the firm's LinkedIn follower count,
mapped on a logarithmic scale to $[0,1]$ so that more prominent employers
receive higher scores. Company names rarely agree with the exact same name across
datasets, so names are first normalized (case, punctuation, legal suffixes such as ``Inc.''), matched exactly when
possible, and otherwise matched when their string similarity exceeds $0.92$,
with some restrictions to prevent issues with distinct names (e.g., ``Company 2''
vs.\ ``Company 22''). In the end, $90$ of the $100$ instance firms matched;
\item the remote share $\rem_j \in [0,1]$ is the fraction of the
pair's postings that allow remote work;
\item the level $\lev_j \in \{0, 1, 2, 3\}$ encodes the posting's stated
seniority: entry, associate, mid--senior, executive/director;
\item the skill profile $\skl_j \in \mathbb{R}^{30}$ records, over LinkedIn's
$30$ skill-function codes (IT, engineering, research, \dots), how often each
code appears among the pair's postings, normalized to unit length.
\end{itemize}
We choose the $m = 100$ suppliers with the largest view counts, while
estimation uses all $709$ filtered suppliers.
 
Individual job seekers are not observable in public data, only per-posting
view and application counts are, so the demand side is modeled through
\emph{types}. A type $t$ is a (level, skill cluster) pair, obtained by
$k$-means clustering of the postings' skill profiles ($3$ clusters, giving
$T = 12$ populated types). Each type's attributes are computed from the
postings in its group, under the maintained assumption that the people who
view and apply to a posting resemble the audience the posting targets:
\begin{itemize}
\item the skill centroid $\skl_t$ is the average of the group's skill profiles,
normalized to unit length (the type's typical mix of job functions); 
\item the level $\lev_t \in \{0,1,2,3\}$ is the group's common seniority level (same
scale as $\lev_j$); 
\item the remote preference $\rem^{\textsf{pref}}_t \in [0,1]$ is the mean remote share of the
group's postings; 
\item the quality index
$\qual_t = 0.7\, \lev_t / \max_t \lev_t + 0.3 \in [0.3, 1]$ is a
candidate-quality score increasing in seniority. Candidates' quality are not observable in the data, so seniority serves as its proxy. We chose an affine map to $[0.3, 1]$. The floor of $0.3$ grants entry-level candidates a
positive baseline appeal rather than zero, while the slope of
$0.7$, combined with the unit coefficient on quality in $\theta_w$,
implies that the most senior candidates are roughly twice as attractive to
employers as the most junior ones. We keep the index within $[0,1]$ so it is on the same scale as the other covariates.
\end{itemize}
Each type's population share (i.e., the probability that a simulated seeker belongs to it) is the
group's share of total posting views, an exposure-based proxy for the
relative size of that seeker segment (views measure attention, not a head
count of individuals).
 
\textbf{Estimation of customers' preference weights, $u$.} On a menu containing a single posting, the MNL
implies that a type-$t$ seeker applies to posting $j$ with probability
$u_{t,j} / (1 + u_{t,j})$, so the \emph{apply odds} (i.e., the probability of
applying divided by the probability of not applying) equal $u_{t,j}$
exactly. Their empirical counterpart in cell $(t,j)$ is
$\applies_{t,j} / (\views_{t,j} - \applies_{t,j})$: applications divided by
non-applying views. We then paremeterize
$u_{t,j} = \exp(\theta_u^\top x_{t,j})$ with
\[
x_{t,j} = (1,\; \sal_j/\median_j \sal_j,\; \ratg_j,\;
\rem^{\textsf{pref}}_t\, \rem_j,\; \langle \skl_t, \skl_j\rangle,\;
-\lvert \lev_t - \lev_j\rvert)
\]
where the first component corresponds to an intercept, the second component is the salary relative to the market median, the third component is the employer rating, the fourth component is a remote interaction (large only when a remote-oriented type meets a remote-offering job), the fifth component is the skill-profile similarity (measured as the cosine between both skill vectors), and a negative seniority gap. This construction makes the log of the odds linear in $\theta_u$. With a half-count correction for empty cells, the empirical
log-odds
$\log(\applies_{t,j} + \tfrac12) - \log(\views_{t,j} - \applies_{t,j} +
\tfrac12)$ of the $457$ populated cells are regressed on $x_{t,j}$ by a 
weighted least squares with weights $\views_{t,j}$. The seniority-gap coefficient is
not identified (populated cells are same-level pairs) and is chosen as a
prior. Robustness checks show
that no coefficient is precisely estimated. Accordingly, $\hat\theta_u$
serves to \emph{calibrate} a realistic instance and should not be interpreted as a
precise behavioral model of job-seeker preferences. The $n = 500$
individuals are then sampled as a random-coefficients mixture: each
individual draws a type according to the population shares and perturbs both
its attributes and its weights. On the attribute side, the skill
profile receives independent Gaussian noise in each of the $30$ coordinates
and is re-normalized to unit length
($\skl_i \propto \skl_t + 0.2\,\xi_i$, $\xi_i \sim \mathcal{N}(0, I)$). The level
is perturbed as
$\lev_i = \mathrm{clip}\bigl(\mathrm{round}(\lev_t + \eta_i)\bigr)$ with
$\eta_i \sim \mathcal{N}(0, 0.35^2)$, so most individuals keep their type's
level and roughly $15\%$ move one unit. The remote preference, quality,
and compensation expectation receive analogous Gaussian noise (standard
deviation $0.2$), clipped back to their ranges. On the weights' side, each
individual independently receives coefficients
$\theta_{u,i} = \hat\theta_u + 0.15\,\varepsilon_i$ with
$\varepsilon_i \sim \mathcal{N}(0, I)$. Finally,
$u_{i,j} = \exp(\theta_{u,i}^\top x_{i,j})$ with covariates evaluated at the
individual's attributes. Perturbing attributes and weights is what
turns the discrete $12$-type mixture into a well-behaved population in the
spirit of mixed MNL: individuals of a common type differ both in what they
\emph{are} (the attributes entering $x_{i,j}$) and in what they \emph{value}
(the weights $\theta_{u,i}$), so the $n$ preference rows are pairwise
distinct yet remain related to the estimated types.

\textbf{Choice of the suppliers' preference weights, $w$.} Supplier weights are computed as
$w_{i,j} = \exp(\theta_w^\top z_{i,j})$ with 
\(
z_{i,j} = (1,\; \qual_i,\; \langle \skl_i, \skl_j\rangle,\;
-\lvert \lev_i - \lev_j\rvert)
\)
where the first component is an intercept, then the candidate's quality index
(increasing in seniority, perturbed per individual), followed by the same skill cosine as
the customer side, and finally the seniority gap. We fix the coefficients
$\theta_w = (-0.3,\, 1.0,\, 1.2,\, 0.6)$: employers prefer higher-quality,
better-matched, similarly-senior candidates. Note that all covariates are observable;
only $\theta_w$ is assumed. Both sides' outside options are normalized to
$1$.

\textbf{Correlation structure of reward values and preference weights.} Let $a_{i,j}$ denote the
\emph{attraction score} of pair $(i,j)$: the value of
$\log(u_{i,j} w_{i,j})$, standardized to mean zero and unit variance across
all pairs. Reward values are drawn as
\[
r_{i,j} = s^{\ast}
\exp\Bigl(\sigma \bigl(\rho\, a_{i,j} +
\sqrt{1 - \rho^{2}}\;\epsilon_{i,j}\bigr)\Bigr),
\qquad \epsilon_{i,j} \sim \mathcal{N}(0,1)\ \text{i.i.d.},
\]
with spread $\sigma = 0.5$ and scale
$s^{\ast} = 0.2 \cdot \median_j \sal_j$, a $20\%$ commission on the median
posted salary. Note that the value 0.2 does not play a role in the results since it is just a scaling parameter. The term in parentheses is a standard normal for \emph{every}
$\rho$, so the distribution of reward values (log-normal with median $s^{\ast}$)
is identical along the range of $\rho$-values, while
$\mathrm{Cor}\bigl(\log r_{i,j},\, \log(u_{i,j} w_{i,j})\bigr) = \rho$ by
construction. We remark that only the noise $\epsilon$ is redrawn per $\rho$, while other random objects $u$, $w$, and the individuals are generated once and reused, so the experiment isolates the correlation effect.

\textbf{Connectivity check.} The resulting market does not decompose into independent
submarkets.

}

\end{document}